\documentclass[review,authoryear]{elsarticle}
\journal{Journal of Econometrics}
\usepackage{xr-hyper}
\RequirePackage[OT1]{fontenc}
\RequirePackage{amsthm,amsmath,amsxtra}
\RequirePackage{natbib}
\setcitestyle{citesep={,}}
\RequirePackage[colorlinks]{hyperref}

\usepackage{enumitem}
\usepackage{colortbl,xcolor,makecell}
\usepackage{array}
\usepackage{mathrsfs}
\usepackage{amssymb,bbm}
\usepackage{amsfonts}
\usepackage{amsmath,amssymb}
\usepackage{booktabs}
\usepackage{graphicx}
\usepackage{graphics}
\usepackage{epsfig}
\usepackage{subfigure}
\usepackage{caption}
\usepackage{amscd}
\usepackage{color}
\usepackage{mathrsfs}
\usepackage{latexsym,bm}
\usepackage{array}
\usepackage{verbatim}
\usepackage{threeparttable}
\usepackage{algorithm}
\usepackage{algpseudocode}
\usepackage{url}

\numberwithin{equation}{section}

\newtheorem{theorem}{Theorem}

\newtheorem{corollary}{Corollary}

\newtheorem{assumption}{Assumption}

\newtheorem{remark}{Remark}

\newtheorem{definition}{Definition}

\DeclareMathOperator{\spa}{sp}
\DeclareMathOperator{\vect}{vec}

\newcommand{\mbf}{\mathbf}
\newcommand{\mbb}{\mathbb}
\newcommand{\mrm}{\mathrm}
\newcommand{\mc}{\mathcal}

\newcommand{\bs}{\boldsymbol}

\newcommand{\R}{\mathbb{R}}
\newcommand{\Pro}{\mathbb{P}}
\newcommand{\E}{{\mathbb{E}}}
\newcommand{\mA}{\mathcal{A}}
\newcommand{\X}{\mathbf{X}}
\newcommand{\Z}{\bm{z}}

\newcommand{\M}{\mathbf{M}}
\newcommand{\Err}{\mathbf{E}}

\newcommand{\Var}{\mrm{Var}}
\newcommand{\Cov}{\mrm{Cov}}

\newcommand{\supp}{\mrm{supp}}
\newcommand{\ad}{\mrm{ad}}

\makeatletter
\def\namedlabel#1#2{\begingroup
	#2%
	\def\@currentlabel{#2}%
	\phantomsection\label{#1}\endgroup
}
\makeatother

\makeatletter
\newcommand*{\addFileDependency}[1]{
	\typeout{(#1)}
	\@addtofilelist{#1}
	\IfFileExists{#1}{}{\typeout{No file #1.}}
}
\newcommand*{\rom}[1]{\expandafter\@slowromancap\romannumeral #1@}
\makeatother

\newtheorem*{assumption*}{\assumptionnumber}
\newcommand{\assumptionnumber}{}

\makeatletter
\newenvironment{assumptionprime}[1]
{%
	\renewcommand{\assumptionnumber}{Assumption #1$'$}%
	\begin{assumption*}%
		\protected@edef\@currentlabel{#1$'$}%
	}
	{%
	\end{assumption*}
}
\makeatother

\makeatletter
\newenvironment{assumptionpprime}[1]
{%
	\renewcommand{\assumptionnumber}{Assumption #1$''$}%
	\begin{assumption*}%
		\protected@edef\@currentlabel{#1$''$}%
	}
	{%
	\end{assumption*}
}
\makeatother

\newcommand*{\myexternaldocument}[1]{%
	\externaldocument{#1}%
	\addFileDependency{#1.tex}%
	\addFileDependency{#1.aux}%
}
\myexternaldocument{Supp9}

\graphicspath{{./fig/}}
\begin{document}
\begin{frontmatter}


\title{Adaptive Change Point Detection with Matrix Time Series: Leveraging Structured Mean Shifts}
\author{Xinyu Zhang}
\ead{xyzhang@sfs.ecnu.edu.cn} 
\affiliation{organization={KLATASDS-MOE, School of Statistics, East China Normal University},
	city={Shanghai},
	postcode={200062}, 
	country={China}} 
\author{Kung-Sik Chan}
\ead{kung-sik-chan@uiowa.edu}  
\affiliation{organization={Department of Statistics and Actuarial Science, University of Iowa},
            city={Iowa City},
            postcode={52246}, 
            state={Iowa},
            country={USA}}

\begin{abstract}

 The components of a high-dimensional time series are often assembled into a matrix to facilitate studying their interrelationship, e.g., in studying traffic dynamics, data may be cross-classified according to pick-up regions (rows) and hours (columns). 
 We focus on detecting mean shifts with unknown change point locations in matrix time series. 
 Such structural changes may exhibit mode-specific alignment, with affected components clustering along certain rows or columns, e.g., changes may only occur over certain hours. Exploiting this structure can substantially improve detection power. 
 We propose an adaptive testing procedure that leverages mode-specific aggregation, while being robust to general, non-aligned changes.  For  inference, we establish the validity of a Gaussian multiplier bootstrap and introduce a computationally efficient parallel bootstrap procedure.
 On the theoretical side, we derive non-asymptotic size and power guarantees under temporal dependence and high dimensionality.
 To obtain the theoretical results, we derive sharper bounds on Gaussian and multiplier bootstrap approximation, which are of independent interest for high dimensional problems with diverging sparsity. 
\end{abstract}


\end{frontmatter}

\section{Introduction}\label{sec:intro}

Structural breaks are ubiquitous in econometrics, for instance, in delineating macroeconomic regimes, studying financial market condition shifts, and tracing evolving  social activity patterns. Detecting such changes is essential for policy analysis, forecasting, and risk management. 
In econometrics, structural change has been studied across various objects of interest, including but not limited to parameter instability in parametric models \citep{andrews1993tests,davis2016consistency}, changes in the mean or level \citep{gorecki2018change}, curve shape  \citep{jiang2020time}, second-order structures \citep{li2023detection}, and factor models  \citep{barigozzi2018simultaneous,ma2018estimation,fu2023testing,bai2024likelihood}.
Among these, changes in the mean (or level) provide a simple and directly interpretable form of structural change. In many applications, policy interventions, market regime shifts, or demand fluctuations manifest as shifts in average outcomes, such as employment levels, consumption, or transaction volumes. This motivates our focusing on mean change as a primary object of interest in this paper.

In many modern econometric applications, the data naturally take the form of matrix-valued time series, where multiple time series are observed across meaningful dimensions such as regions, sectors, or time intervals. Examples include panels of macroeconomic indicators (e.g., GDP, CPI, and interest rates) across countries, financial characteristics across firms, and spatial-temporal measures such as transportation demand across locations and time of day. In these settings, the matrix representation captures structured inter-series relationships, with observations within the same row or column often exhibiting stronger dependence due to shared economic or behavioral factors.
Recent work has developed models tailored to matrix-valued time series, including matrix autoregressive models \citep{chen2021autoregressive,chan2025large} and matrix factor models \citep{chen2023statistical,chen2022factor,wang2019factor,yu2022projected,chang2023modelling,he2023oneway}. 
However, structural change in matrix-valued time series remains largely underexplored.

\begin{figure}[htbp]
  \centering
  \includegraphics[width=0.45\textwidth]{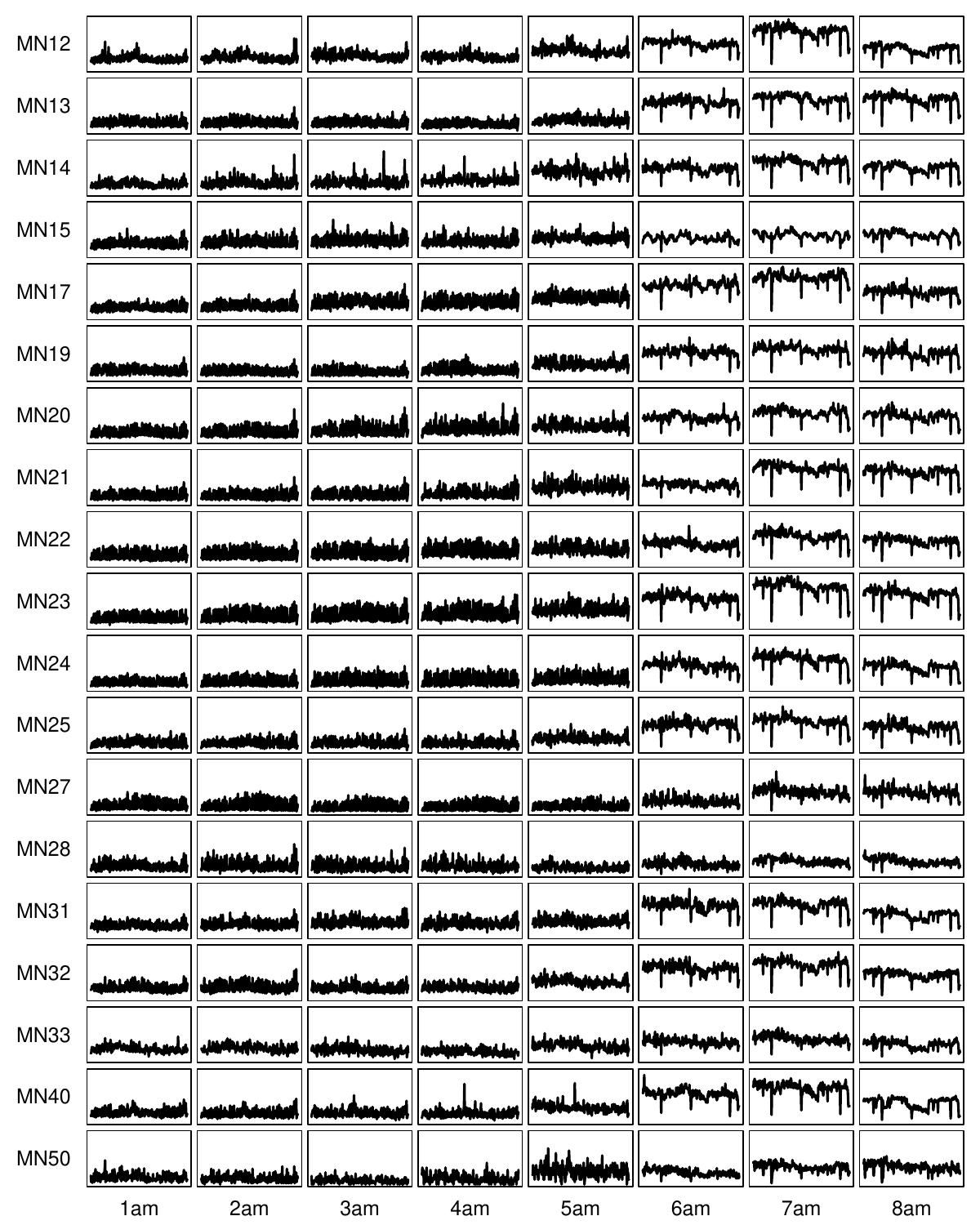}
  \includegraphics[width=0.45\textwidth]{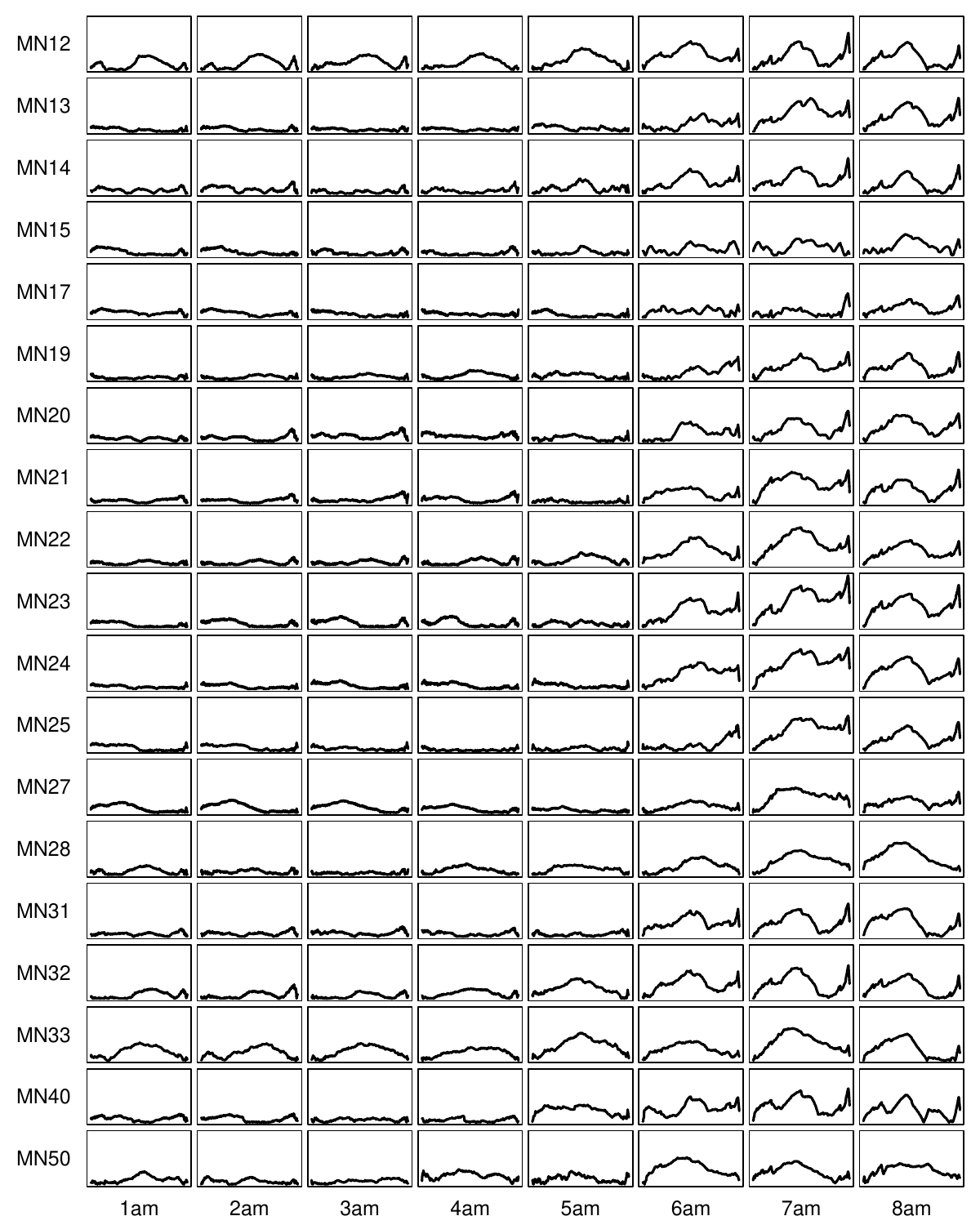}
  \caption{Time series plot (left) and CUSUMs (right) for taxi pickup counts at each zones (row) within each hour (column) in Manhattan, New York  city, in 2017.}
  \label{fig:cusum}
\end{figure}

Motivated by these considerations, we study the problem of testing whether the mean function of a matrix-valued time series remains constant over time against the alternative that it is piecewise constant. 
A key challenge in matrix data is that structural breaks may occur in a coordinated manner across related series. For example, a macroeconomic shock may affect multiple countries simultaneously, while changes in commuting patterns may primarily impact specific hours across locations. This suggests that structural changes may align along rows or columns of the matrix. 
For illustration, consider the matrix of taxi pickup counts across locations and hours of the day (see Section \ref{sec:app} for more details and analysis). 
Figure \ref{fig:cusum}  displays the matrix time series plots and their corresponding  CUSUM curves.  These plots suggest that structural breaks occurred sporadically,  with apparent alignment patterns, such as column-wise changes during morning rush hours (e.g., 6-8 am) and row-wise changes in specific locations (e.g., MN12 and MN33).
Such patterns highlight the importance of incorporating matrix structure in change point analysis and suggest that pooling information along relevant dimensions can improve detection and interpretability.

From a methodological perspective, change point detection has been extensively studied in univariate and multivariate settings, with the CUSUM approach playing an important role \citep{csorgo1997limit}. In high-dimensional settings, additional challenges arise due to cross-sectional dependence and the possibility that structural changes affect only a subset of the series. Various aggregation methods have been proposed to address these issues, such as $l_2$-aggregation \citep{horvath2012changepoint,enikeeva2019highdimensional,wang2018changepoint}, $l_{\infty}$-aggregation \citep{jirak2015uniform,yu2021finite}, thresholded $l_1$-aggregation \citep{cho2015multiplechangepoint}, the double CUSUM approach  \citep{cho2016changepointa},  projection-based methods \citep{wang2018high}, and adaptive approaches
\citep{liu2020unified,zhang2022adaptive,wang2023computationally}.
For matrix data, a natural approach is to vectorize the matrix observations and apply existing methods.
However, this ignores the row–column structure and associated dependence patterns, potentially resulting in suboptimal performance \citep{chen2023statistical}, reduced power and limited interpretability. 

Motivated by the alignment patterns discussed above, we propose mode-specific test statistics that aggregate CUSUM signals along rows and columns, allowing for more effective detection of structured changes in matrix-valued time series.
The key idea is to first compute element-wise CUSUM statistics and then combine them within each row or column to capture coordinated changes across related series. 
When a structural break occurs simultaneously across multiple entries in a row or column, such aggregation reinforces the common signal while reducing noise. The resulting statistic takes the maximum over rows or columns to identify the most prominent change pattern.
This approach provides a powerful and interpretable tool for detecting and characterizing mode-specific mean changes in matrix-valued time series.

While the proposed mode-specific tests are effective for detecting changes aligned along rows or columns, structural changes in practice may exhibit more complex patterns. In particular, changes may be dispersed across the matrix or affect only a small subset of entries, making the alignment less pronounced.
To accommodate such scenarios, we extend the proposed framework by developing an adaptive testing procedure that remains effective across a wide range of change patterns. The idea is to combine multiple test statistics, each targeting different types of structural changes, so that the procedure can automatically adjust to the underlying signal structure. The combination is implemented via p-value aggregation, allowing the test to retain good power whether the changes are aligned, dispersed, or sparse.

To conduct inference, it is necessary to approximate the distribution of the proposed statistics. In high-dimensional settings with temporal and cross-sectional dependence, this distribution is generally intractable, and classical extreme value approximations may converge slowly \citep{csorgo1997limit}.
We therefore adopt a dependent Gaussian multiplier bootstrap approach, which provides accurate approximation by capturing the dependence structure of the data. 
On the theoretical side, we establish refined approximation results that improve existing bounds in terms of sparsity.
This improvement is crucial in our setting, where the sparsity parameter may grow with the dimension. Our sharper bound is also of independent interest for other high-dimensional problems involving diverging sparsity and temporal dependence.
For the adaptive test, a parallel bootstrap is proposed with manageable computational cost and robust performance.


Our contribution is twofold.
First, we study mean change point testing for matrix-valued time series, a setting that has received limited attention in the change point literature. 
We develop a framework based on mode-specific test statistics that exploit row–column alignment patterns, together with an adaptive procedure that combines multiple tests to handle a wide range of change structures. 
Unlike existing methods designed for vector-valued data, which often rely on temporal independence, our approach accommodates both temporal and cross-sectional dependence while leveraging the matrix structure to detect and interpret structured changes. 
From an econometric perspective, the proposed framework addresses several key challenges, including detecting structural breaks in high-dimensional dependent systems, accommodating joint temporal and cross-sectional dependence, and exploiting structured interactions across economic units through the matrix formulation.
Second, on the theoretical side, we establish sharper Gaussian and bootstrap approximation bounds for dependent data, relative to \cite{chang2024central}, with improvements in the order of the sparsity parameter. These results lead to more accurate distributional approximations for the proposed test statistics. They are also relevant for a broader class of high-dimensional problems where signals are aggregated over groups or clusters of variables, such as testing for differences in covariance structures across submatrices and detecting best subgroup effects where only a subset of units exhibits a pronounced signal.


The rest of this paper is organized as follows.
In Section \ref{sec:method}, we elaborate the matrix change point detection problem, and propose mode-specific tests as well as an adaptive test.
Section \ref{sec:GA} introduces the Gaussian and bootstrap approximation, derives an  improved approximation bound, and elaborates their implementation for our proposed tests.
Section \ref{sec:theory} provides the theoretical size validity and power results for the proposed  tests.
Sections \ref{sec:simu} and \ref{sec:app} provide numerical illustrations including extensive simulations and a real application.
Section \ref{sec:dis} concludes and provides discussion.
The \emph{Supplementary Materials} include all  proofs and additional numerical results.

\section{The matrix change point test statistic}\label{sec:method}

In this section, we develop testing procedures for matrix-valued time series. We first introduce mode-specific tests that exploit row–column alignment structures, and then extend them to an adaptive procedure that accommodates a broader range of change patterns.
\subsection{A mode-specific test}\label{sec:modetest}
Consider a sequence of 
matrix time series $\X= \{\X_1,\ldots,\X_N \}$ with $\X_i \in \R^{p_1\times p_2}$ and the following mean shift model:
\begin{equation}\label{eq:model}
	\X_i=\M+\bs{\delta}_N \mbb{I}(i>u)+\Err_i,
\end{equation}
where $\M\in \R^{p_1\times p_2}$ is the baseline matrix mean,  $\bs{\delta}_N \in \R^{p_1\times p_2}$ is the matrix mean-shift at the unknown epoch (location) $1<u<N$,
and $\Err_i \in \R^{p_1\times p_2}$ are stationary $\alpha$-mixing time series with zero mean; see \eqref{eq:mixing} for the definition $\alpha$-mixing coefficients.
WLOG, assume $\M=\bm{0}$.

Our goal is to test the existence of a change point, i.e.,
\begin{equation*}
	H_0: \bs{\delta}_N=0 \,\,\text{versus}\,\,  
	H_1: \exists 1 \leq u \leq N-1 \,\,\text{and}\,\, \bs{\delta}_N \neq 0. 
\end{equation*}
We generalize the well-known cumulative sum (CUSUM) to define the CUSUM matrix $\mbf{C}_{n}(\X) \in \R^{p_1\times p_2}$ with $1\leq n \leq N-1$ as follows:
\begin{align}\label{eq:cn}
	\begin{split}
		\mbf{C}_{n}(\X) = &\sqrt{\frac{n(N-n)}{N}} \left(\frac{1}{N-n}\sum_{i=n+1}^N \X_{i}-
		\frac{1}{n}\sum_{i=1}^n \X_{i}\right).
	\end{split}
\end{align}

For matrix time series data, the challenge is how to aggregate the CUSUM matrix for effecting efficient change point detection.
As shown in Figure \ref{fig:cusum}, for matrix time series, the change point   structure may be concentrated  within certain rows or columns.
To leverage mode-specific changes, we propose 
to aggregate the CUSUM matrix using the following mode-specific norms.
Specifically, given a matrix $\mbf{A} \in \R^{p_1\times p_2}$, we define $\|\mbf{A}\|_{[mode,q]}$ with $mode=1,2$ and $q=2$ (representing $l_2$ norm) as
\begin{align}\label{eq:mode2}
  \|\mbf{A}\|_{[1,2]} =\max_{1\leq j_1 \leq p_1}(\sum_{j_2=1}^{p_2} |a_{j_1,j_2}|^2)^{1/2}, \,\,\,
  \|\mbf{A}\|_{[2,2]} =\max_{1\leq j_2 \leq p_2}(\sum_{j_1=1}^{p_1} |a_{j_1,j_2}|^2)^{1/2}.
\end{align}
Note that both are matrix norms, since they are exactly the induced $\|\cdot\|_{2,\infty}$ and $\|\cdot\|_{1,2}$ matrix norm, respectively \citep[Section 2.3]{golub2013matrix}. 
We nevertheless retain the $[mode,q]$ notation to facilitate a unified treatment when extending to other choices of $[mode,q]$ later.
Then, the test statistics $T_{[mode,2]}(\X)$ are given by 
\begin{align}\label{eq:T}
  T_{[1,2]}(\X)=\max_{\nu \leq n \leq N-\nu}\|\mbf{C}_{n}(\X)\|_{[1,2]},\,\,\,
  T_{[2,2]}(\X)=\max_{\nu \leq n \leq N-\nu}\|\mbf{C}_{n}(\X)\|_{[2,2]},
\end{align}
where $1< \nu < N/2$ is a boundary removal parameter and possibly growing to infinity as $N \to \infty$.
Note that the test $T_{[1,2]}(\X)$ ($T_{[2,2]}(\X)$) is  sensitive to concentration of change points within specific rows (columns).

\begin{remark}[Motivation of the test statistic]\label{rem:lr}
	The test statistic $T_{[1,2]}(\X)$ and $T_{[2,2]}(\X)$ can be motivated by the generalized likelihood ratio test under certain circumstance. 
	Take $T_{[1,2]}(\X)$ as the example.
	Consider the situation that the changes take place at time $n$ and the $j$th row is the only row where changes may occur, i.e.,  
	$[\bs{\delta}_{N}]_{-j:}$, denoted as all other rows except for the $j$th row of $\bs{\delta}_{N}$, is 0.
	Denote $[\mbf{C}_{n}(\X)]_{ j:}$ as the $j$th row of the CUSUM matrix.
	Note that $T_{[1,2]}(\X)=\max_{\nu \leq n \leq N-\nu}\max_{1\leq j \leq p_1}\|[\mbf{C}_{n}(\X)]_{ j:}\|_2$.
	Consider a simplified case where each element in the $\Err_i$ matrix is i.i.d and follows $\mc{N}(0,\sigma^2) $ for some $\sigma>0$. Specifically, denote $\mc L(\M,\bs{\delta}_{N}, \sigma^2;\X)$ as the likelihood w.r.t the parameter $\M$,  $\bs{\delta}_{N}$ and $ \sigma^2$ given observations $\X$.
	Then,
	\begin{align*}
		\log \frac{\max_{\M,\bs{\delta}_{N}\in \R^{p_1\times p_2},[\bs{\delta}_{N}]_{ -j:} = 0, \sigma^2\in \R} \mc L(\M,\bs{\delta}_{N}, \sigma^2;\X)}{\max_{\M\in \R^{p_1\times p_2},\sigma^2\in \R} \mc L(\M,\mathbf{0}_{p_1\times p_2}, \sigma^2;\X)}
		=-\frac{N}{2}\log\left(1-\frac{\|[\mbf{C}_{n}(\X)]_{ j:}\|_2^2}{RMS}\right).
	\end{align*}
	where $RMS=\sum_{i=1}^N \|\X_i-\bar{\X}\|_2^2$. Clearly, the likelihood ratio statistic is an increasing function of  ${\|[\mbf{C}_{n}(\X)]_{ j:}\|_2^2}/{RMS}$. Thus, the likelihood ratio test statistic for unknown row mean shift at unkown time is an increasing function of  $T_{[1,2]}(\X)/RMS$ hence the  choice of $T_{[1,2]}(\X)$  for detecting unknown row mean shift.
\end{remark}

We propose to assess the statistical significance of $T_{[mode,2]}(\X)$ via bootstrap. 
Let $T_{[mode,2]}^b(\X), b=1,\cdots, B$ be $B$ independent bootstraps.
The specific bootstrap procedures we use will be elaborated separately in Section \ref{sec:bootstrap}; see (\ref{eq:tmodeqb}). The distribution of $T_{[mode,2]}(\X)$ is then approximated by the empirical distribution of $T_{[mode,2]}^b(\X)$. 
For any $\alpha \in (0,1)$, denote the $1-\alpha$ empirical bootstrap quantile  by
\begin{equation*}
	\hat{d}_{1-\alpha,[mode,2]}\equiv
	\inf\left\{t \in \R: \frac{1}{B}\sum_{b=1}^B \mbb{I}\left(T_{[mode,2]}^b(\X) \leq t\right) \geq 1-\alpha\right\},\, mode \in\{1,2\}.
\end{equation*}
Then, if $T_{[mode,2]}(\X) \geq \hat{d}_{1-\alpha,[mode,2]}$, it would indicate that $H_0$ is rejected at significance level $\alpha$.
Alternatively, $H_0$ is rejected if and only if the bootstrap p-value  $\hat{P}_{[mode,2]} \leq \alpha$, where 
\begin{equation}\label{eq:pmode2}
	\hat{P}_{[mode,2]}=\frac{1}{B}\sum_{b=1}^B \mbb{I}\left(T_{[mode,2]}^b(\X) > T_{[mode,2]}(\X)\right).
\end{equation}

Although the test statistics in (\ref{eq:T}) are motivated by mode-specific break alignments (as illustrated in Figure \ref{fig:cusum}), they remain theoretically valid regardless of whether such alignment is present; see Section \ref{sec:theory}.
To further enhance performance across more general scenarios, we next develop an adaptive procedure that accommodates a wider range of break structures beyond mode-specific changes.

\subsection{An adaptive test}\label{sec:adtest}

In addition to aligned change patterns, structural changes may also exhibit more dispersed or heterogeneous behavior across the matrix. To accommodate such scenarios, it is useful to consider aggregation schemes that are not tied to specific row or column structures. We therefore extend the $[mode,2]$ norms by introducing a non-mode-specific aggregation that operates on the vectorized matrix.
Specifically, we define
\begin{equation}\label{eq:dot2}
	\|\mbf{A}\|_{[\cdot,2]} =(\sum_{j=1}^{s} |\vect(\mbf{A})_{(j)}|^2)^{1/2},
\end{equation}
where $|\vect(\mbf{A})_{(1)}| \geq \ldots \geq |\vect(\mbf{A})_{(s)}| $ is the order statistics of the absolute values in descending order with some $1<s<p$.
Setting $mode=\cdot$ signifies that it is non-mode-specific or mode-free that is computed from the vectorized matrix.

Despite the superficial difference between the $[mode,2]$ norms with $mode=1,2$ in (\ref{eq:mode2}) and that with $mode=\cdot$ in (\ref{eq:dot2}), they can be unified as follows.
The $[\cdot,2]$ norm computes the $l_2$ norm of the leading $s$ elements (in absolute value), i.e., the maximum of the $l_2$ norm of all subvectors of $\vect(\mbf{A})$ with $s$ elements.
In comparison, the $[1,2]$ and $[2,2]$ norms are constrained versions of $[\cdot,2]$ norm: the $[1,2]$ norm ($[2,2]$ norm) is the maximum of the $l_2$ norm of all subvectors of $\vect(\mbf{A})$ with $p_2$ elements ($p_1$ elements)
and align in the same row (column).
For the choice of $s$, 
we restrict $s \leq \max(p_1,p_2)$ to balance flexibility and stability in high-dimensional settings.

In addition, we consider scenarios where signals are highly concentrated in a small number of entries. To capture such patterns, we incorporate the maximum norm of the CUSUM matrix. Under our notation, this corresponds to replacing the $l_2$ norm with the $l_{\infty}$ norm for any $mode \in \{1,2,\cdot\}$, i.e.,
\[
\|\mbf{A}\|_{\max}=\|\mbf{A}\|_{[1,\infty]}=\|\mbf{A}\|_{[2,\infty]}=\|\mbf{A}\|_{[\cdot,\infty]}.
\]

Similar to (\ref{eq:T}), for $mode \in \{1,2,\cdot\}$ and $q \in \{2,\infty\}$, define 
\begin{align}\label{eq:Tmodeq}
	\begin{split}
		T_{[mode,q]}(\X)&=\max_{\nu \leq n \leq N-\nu}\|\mbf{C}_{n}(\X)\|_{[mode,q]}
	\end{split}
\end{align}
where $\nu$ is a boundary removal parameter.
Estimate the p-value $\hat{P}_{[mode,q]}$ for each test as in (\ref{eq:pmode2}), and then combine these individual tests using the Tippett's method \citep{tippett1931methods}, i.e., use the minimum of p-values as a test statistic to form an adaptive test.
Specifically,
\begin{align}\label{eq:Tad}
	T_{\ad}(\X)=\min_{mode \in \{1,2,\cdot\}, \, q \in \{2,\infty\}}\hat{P}_{[mode,q]}.
\end{align}
Note that each $T_{[mode,q]}(\X)$ demonstrates different power performance against specific alignment alternatives.
In general, $T_{[1,2]}(\X)$ (or $T_{[2,2]}(\X)$) is sensitive to alternatives with change point  concentration within specific rows (or columns), while $T_{[\cdot,2]}(\X)$ gains  power when such matrix structure is less pronounced or absent.
And the test with $q=\infty$ complements the case where the signal shows a highly  concentrated alignment pattern.
By combining these statistics, the adaptive test can automatically adjust to the underlying signal structure without prior knowledge of the alignment pattern.

To approximate the distribution of the test statistic $T_{\ad}(\X)$, in Section \ref{sec:bootstrap}, we propose a new method called parallel bootstrap to implement the test.

\begin{remark}[Adaptive versus mode-specific tests]\label{rem:vs}
In practice, the adaptive test can be viewed as a robust default choice, as it maintains good power across a wide range of alternatives, including both mode-specific and more general changes. When prior knowledge suggests that changes are aligned along a specific dimension (e.g., rows or columns), the corresponding mode-specific tests may be preferred due to their higher sensitivity in such settings. 
This reflects a tradeoff between robustness and efficiency that is common in high-dimensional testing problems.
\end{remark}

\section{Gaussian and bootstrap approximations }\label{sec:GA}

In this section, we develop inference procedures for the proposed test statistics by constructing bootstrap-based critical values. We begin in Section \ref{sec:approximation} by reviewing the Gaussian and bootstrap approximation framework, and in Section \ref{sec:spa} we establish refined theoretical bounds. Based on these results, Section \ref{sec:bootstrap} introduces implementable bootstrap procedures, including the Gaussian multiplier bootstrap for the mode-specific tests and a parallel bootstrap for the adaptive test.

\subsection{Introduction of the Gaussian and bootstrap approximation}\label{sec:approximation}

Gaussian and bootstrap approximations have become important tools for inference in high-dimensional settings, particularly for statistics involving maxima or other nonlinear functionals, such as those arising in change-point problems \citep{liu2020unified,yu2021finite} and multiple testing \citep{chang2017testing}.
A seminal contribution is \citet{chernozhukov2013gaussian}, which initiated a line of research on Gaussian approximation for sums of high-dimensional independent random vectors; see \citet{chernozhukov2023highdimensional} and the reference therein for a comprehensive overview.
Extensions to dependent data have been developed in \citet{zhang2017gaussian,zhang2018gaussian,chernozhukov2019inference,mies2023sequential}, and more recently, \citet{chang2024central} provided a unified treatment under temporal dependence.
In this section, we review Gaussian and bootstrap approximation results from \cite{chang2024central} and present them in a form suitable for analyzing the proposed test statistics.

To formalize the approximation, consider a sequence of centered random vectors $\Z_1,\dots,\Z_N \in \R^p$ with $\Z_i=(z_{i1},\dots,z_{i p})^{\top}$, which may exhibit temporal dependence.
Consider the sum $$S^z_N\equiv(S^Z_{N1},\dots,S^Z_{Np})^{\top}\equiv\frac{1}{\sqrt{N}}\sum_{i=1}^N \Z_i$$ 
with the long-run covariance matrix  $\bm{\Xi}=\Cov(S^z_N)$.
Under suitable conditions, the distribution of the sum $S^z_N$
can be approximated by that of a Gaussian random variable $\bm{G} \sim \mc{N}(0, \bm{\Xi})$.
To quantify the accuracy of this approximation, consider the Kolmogorov distance $\rho_N(\mA)\equiv\sup_{A \in \mA}|\Pro(S^z_N\in A)-\Pro(\bm{G}\in A)|$,
where  $\mA$ is a class of Borel sets in $\R^p$.

In practice, the covariance matrix $\bm{\Xi}$ is unknown, making direct Gaussian approximation infeasible. To address this, a bootstrap approach is adopted, based on a kernel estimator of $\bm{\Xi}$.
 Let $\mc{K}$ be a symmetric kernel function that is continuous at 0 with $\mc{K}(0)=1$, and let $l_N>0$ be the bandwidth.
Let $\bar{\Z}\equiv\frac{1}{{N}}\sum_{i=1}^N \Z_i$, and 
\begin{equation}\label{eq:xihat}
	\widehat{\bm{\Xi}}_N=\sum_{j=-N+1}^{N-1}\mc{K}(j/l_N)\hat{\bm{\Gamma}}_j
\end{equation}
  with the sample autocovariances $\hat{\bm{\Gamma}}_j=N^{-1}\sum_{t=j+1}^N(\Z_{t}-\bar{\Z})(\Z_{t-j}-\bar{\Z})^{\top}$ for $j \geq 0$ and $\hat{\bm{\Gamma}}_j=\hat{\bm{\Gamma}}_{-j}^{\top}$ for negative $j$.
Consider the multiplier bootstrap approach with dependent multipliers $(e_1,\dots,e_N)\sim \mc{N}(0, \bm{\Theta})$, independent of $ \{ \Z_{1},\dots,\Z_{N} \}$, where $\bm{\Theta}=(\theta_{i,j})_{N \times N}$ is such that $\theta_{i,j}=\mc{K}((i-j)/l_N)$. 
The rationale is that conditional on the observations $\Z$, the sum $$S_N^{ez}\equiv\frac{1}{\sqrt{N}}\sum_{i=1}^N e_i(\Z_i-\bar{\Z}) \sim \mc{N}(0,\widehat{\bm{\Xi}}_N)$$
is a Gaussian-to-Gaussian approximation of $\bm{G}$, respectively with covariance matrix $\widehat{\bm{\Xi}}_N$ and $\bm{\Xi}$.
The Kolmogorov distance $\rho_N^{B}(\mA)\equiv\sup_{A \in \mA}|\Pro(S_N^{ez}\in A \mid \Z)-\Pro(\bm{G}\in A)|$ measures the accuracy of the bootstrap approximation over $\mA$.

\begin{remark}[Approximation for independent data]
When $\Z_i$ are independent, the long-run covariance reduces to the covariance, and the bootstrap can be simplified by using independent multipliers, i.e., $\bm{\Theta}$ reduces to the identity matrix. In this case, the bootstrap statistic remains Gaussian with covariance given by the sample covariance matrix. See \citet{chernozhukov2023highdimensional} for further discussion.
	
\end{remark}

To analyze the distribution of the proposed test statistics, which involve maxima over structured subsets of coordinates, we consider a class of sparsely convex sets introduced by \citet{chernozhukov2017central}. This class is well suited for high-dimensional problems with sparse or structured signals and aligns naturally with the $[mode,q]$ norms used in our change point detection framework.
In our setting, we further refine this class by incorporating structural constraints on the support.
Specifically, for any vector $\bm{w}=(w_1,\cdots, w_p)^\top \in \R^p$,  its support is $\supp(\bm{w})=\{1\le j\le p: w_j\not = 0\}$. 
Similarly, for a set $A$ and the indicator function $\mbb{I}(\cdot\in A)$, let $\supp(\mbb{I}(\cdot\in A))$ comprise the collection of all active arguments of the function $\mbb{I}(\cdot\in A)$. 
That is,  $\supp(\mbb{I}(\cdot\in A))=\{1\le i\le p: \exists w, w' \mbox{ with } w_i\not = w'_i, w_j=w'_j,  \forall j\not = i, \mbox{ such that } \mbb{I}(w\in A)\not = \mbb{I}(w'\in A)\}$. 
For some $1 \leq s \leq p$, let $\mc{I}_0$ be the collection encompassing all subsets of $\{1,2,\ldots,p\}$ with at most $s$ elements. 
Now, consider a support collection $\mc{I} \subseteq \mc{I}_0$ and define the generalized sparsely convex sets as follows.

\begin{definition}[Generalized sparsely convex sets]\label{def:spa}
	For integer $s>0$,  $A\subset\R^p$ is said to be an $s$-sparsely convex set if there exists an integer $Q>0$ and convex sets $A_q \subset \R^p, q=1,\dots,Q$, such that $A=\cap_{q=1}^Q A_q$, and the indicator function of each $A_q$, $w\mapsto \mbb{I}(w\in A_q)$ depends  on at most $s$ elements of its argument $w=(w_1,\dots,w_p)$ (each of which is referred to as a main component of $A_q$), i.e.,  $\supp(\mbb{I}(\cdot\in A_q)) \in \mc{I}_0$.
	If, in addition, the main components are further constrained to lie in a given collection such that $\supp(\mbb{I}(\cdot\in A_q)) \in \mc{I}\subseteq \mc{I}_0$, 
    then $A$ is called a generalized $s$-sparsely convex set on $\mathcal{I}$. We denote the class of such sets by $\mathcal{A}^{\mathrm{spa}}(s,\mathcal{I})$.
\end{definition}

The class $\mathcal{A}^{\mathrm{spa}}(s,\mathcal{I})$ extends the standard sparsely convex sets by allowing additional structure in the support, which is useful for 
handling the complexity introduced by maximizing CUSUM statistics over time indices.
Importantly, this extension does not introduce additional technical difficulty in the theoretical analysis.


\subsection{Improved approximation bounds for sparsely convex sets}\label{sec:spa}

In this section, we present the Gaussian and bootstrap approximation bounds on the probability that the sum of $\alpha$-mixing centered  high-dimensional random vectors lie within generalized  sparsely convex sets. 
These results build on \cite{chang2024central}
and refine their bounds  by improving their dependence on the sparsity parameter $s$.
See Remark \ref{rem:improve} for an explanation of how this improvement is achieved.
These results are instrumental in establishing the validity of our change-point tests in Section~\ref{sec:theory}, and are of independent interest in high-dimensional settings with diverging  sparsity level.

We first present some regularity conditions.
Denote by $\mathcal{F}_{-\infty}^t$ and $\mathcal{F}_{t}^\infty$ the $\sigma$-fields generated respectively by $\{\Z_i\}_{i \leq t}$ and $\{\Z_i\}_{i \geq t}$, and
define the $\alpha$-mixing coefficient at lag $k$ as
\begin{equation}\label{eq:mixing}
	\alpha_N(k)\equiv\sup_t\sup_{A\in\mathcal{F}_{-\infty}^t, B\in\mathcal{F}_{t+k}^\infty}|\mathbb{P}(A\cap B)-\mathbb{P}(A)\mathbb{P}(B)|.
\end{equation}
Let $b>0$ be a universal constant and let $B_N\geq 1$ be a sequence of constants which could diverge as $N$ increases.
For the data $\{\Z_i\}$ and the set $\mA_{\spa}(s,\mc{I})$, assume the following conditions hold.

\begin{itemize}
	\item [\namedlabel{mixing}{(Mixing)}]
	For any $k\geq 1$, $
	\alpha_N(k) \leq K_1 \exp(-K_2 k^{\gamma}),
	$ where $K_1>0 ,K_2>0, \gamma>0$ are some universal constants independent of $N$.
	\item [\namedlabel{espa}{(Subexp)}] $\E[\exp(|\Z_{ij}|/B_N)] \leq 2$
	for  all $i=1,\ldots,N$ and $j=1,\ldots, p$.
	\item [\namedlabel{mspad}{(Moment)}] $N^{-1} \Var[\sum_{i=1}^{N}(\bm{v}^{\top}\Z_{i})] \geq b^2$  
	for all $\bm{v}\in\mathbb{S}^{p-1}$ with $\supp(\bm{v}) \in \mc{I}$.
\end{itemize}

Condition~\ref{mixing} assumes geometric $\alpha$-mixing.
Condition~\ref{espa} assumes sub-exponential tails, which are commonly adopted in the high-dimensional statistics literature. This assumption can, in principle, be relaxed to polynomial moment conditions as in \citet{chernozhukov2017central}. For dependent data, \citet{chang2024central} notes that such an extension is feasible but would follow similar lines of argument and is therefore not pursued. For brevity and comparability with the existing dependent-data framework, we focus on the sub-exponential case.
Condition~\ref{mspad} ensures a non-degeneracy condition for sparse projections, which is required for Gaussian approximation and anti-concentration arguments \citep{nazarov2003maximal}.

We now present the Gaussian and bootstrap approximation bounds. 
\begin{theorem} \label{thm:gaussian-dep}
	Assume that Conditions \ref{mixing}, \ref{espa} and \ref{mspad} are satisfied,
	 and $p \geq N^{\tau}$ for some universal constant $\tau>0$.
	Then, it holds that 
	$$\rho_N(\mA_{\spa}(s,\mc{I}))\lesssim \frac{B_{N}^{2/3}s^{(1+4\gamma)/(3\gamma)}(\log p)^{(1+2\gamma)/(3\gamma)}}{N^{1/9}}+\frac{B_{N}s^{13/6}(\log p)^{7/6}}{N^{1/9}}$$
	provided that $(s\log p)^{3-\gamma}=o(N^{\gamma/3})$.
\end{theorem}

\begin{theorem}\label{thm:boot-dep}
	Assume $p \geq N^{\tau}$ for some universal constant $\tau>0$.
	Assume that Conditions \ref{espa} and \ref{mspad} are satisfied.
	Let $\Delta_{N}=\|\widehat{\bm{\Xi}}_{N}-\bm{\Xi}\|_{\max}$.
	Then, it holds that
	$$\rho_N^{B}(\mA_{\spa}(s,\mc{I}))\lesssim s\Delta_{N}^{1/3}(\log p)^{2/3}+\{B_{N}+(s\log p)^{1/2}\}N^{-1}.$$
	
\end{theorem}

The validity of the bootstrap depends on the error of the estimated long run covariance matrix $\Delta_{N}$, which can be quantified under certain conditions. 
Applying Theorems~\ref{thm:gaussian-dep} and~\ref{thm:boot-dep}, we obtain
 $$\sup_{A \in \mA_{\spa}(s,\mc{I})}|\Pro(S_N^{ez}\in A \mid \Z)-\Pro(S_N^z\in A)| \leq \rho_N(\mA_{\spa}(s,\mc{I}))+\rho_N^{B}(\mA_{\spa}(s,\mc{I})).$$

The following remark summarizes the theoretical improvement in our approximation error bounds stated in Theorems \ref{thm:gaussian-dep} and \ref{thm:boot-dep} with respect to the sparsity parameter $s$, explains how this improvement is achieved, and highlights its significance.

\begin{remark}[Theoretical improvement on the bounds]\label{rem:improve}
	Compared to Theorem 7 in \cite{chang2024central}, which establishes a Gaussian approximation bound $ \frac{(B_{N}s)^{2/3}(s^2\log p)^{(1+2\gamma)/(3\gamma)}}{N^{1/9}}+\frac{(B_{N}s)(s^2\log p)^{7/6}}{N^{1/9}}$, our bounds in Theorem \ref{thm:gaussian-dep} improve the dependence on the sparsity parameter $s$ by replacing terms of the form $(s^2 \log p)$ with $(s \log p)$. This leads to sharper rates when $s$ grows with the dimension.
    
    
	The key reason for this improvement lies in a more efficient approximation of sparsely convex sets by convex polytopes using polar duality.
    The complexity of this approximation is measured by a log term of the number of facets of the polytope, which scales as $(pN)^{s^2}$ in \citet{chang2024central} and \citet{chernozhukov2017central}, and therefore contributes to the $s^2$ factor in the bounds. 
    By exploiting a refined geometric construction based on the polar dual construction of polytopes, we reduce this complexity to $(pN)^s$, thereby improving the dependence on $s$.

	
    This improvement has direct implications for the admissible dimensional regime. The approximation error typically depends on both $s$ and $\log p$, and when $s$ diverges with $p$, the sparsity term becomes the dominant factor. In this case, reducing the power of $s$ weakens the rate restrictions linking $p$ and $N$, and hence allows $p$ to grow faster with $N$ within the polynomial-growth regime covered by our theory.
    This technique also applies to independent data, yielding improved rates in terms of the sparsity parameter $s$.

\end{remark}

\subsection{Bootstrap procedures for the matrix change point tests}\label{sec:bootstrap}

In this section we explain how Gaussian and bootstrap approximation introduced above apply to the proposed change point statistics. Recall that the CUSUM statistic can be written as a normalized sum $\mbf{C}_{n}(\X) =\frac{1}{\sqrt{N}}\sum_{i=1}^N \kappa_{n i} \X_{i}$,
where the weights are
\begin{equation}\label{eq:kappa}
	\kappa_{n i}\equiv-\sqrt{\frac{N-n}{n}} \text{ for }1 \leq i \leq n,\,\text{ and }
	\kappa_{n i}\equiv\sqrt{\frac{n}{N-n}} \text{ for } n+1 \leq i \leq N.
\end{equation}
Stacking the CUSUM sequence $(\mbf{C}_{\nu}(\X), \ldots, \mbf{C}_{N-\nu}(\X))$ and vectorizing, we obtain a high-dimensional sum of the form $S_N^z = N^{-1/2}\sum_{i=1}^N \Z_i$ with
\begin{align}\label{eq:zi1}
		\Z_i &=\left(\kappa_{\nu i} \vect(\X_{i}),\ldots,\kappa_{(N-\nu) i} \vect(\X_{i})\right) \\
		&=\left(\kappa_{\nu i}\X_{i11},\ldots,\kappa_{\nu i}\X_{ip_1p_2},\ldots,\kappa_{(N-\nu)i}\X_{i11},\ldots,\kappa_{(N-\nu) i}\X_{ip_1p_2}\right)
		\in \R^{p(N-2\nu+1)},\nonumber
\end{align}
with $p=p_1p_2$.
For a given choice of $[mode,q]$, the distribution function of the test statistic $T_{[mode,q]}(\X) =\max_{\nu \leq n \leq N-\nu}\|\mbf{C}_{n}(\X)\|_{[mode,q]}$ can be written as
\begin{equation}\label{eq:lieinset}
	\Pro(T_{[mode,q]}(\X)\leq t)=\Pro(S^z_N \in A)
\end{equation}
where $A=\cap_{J \in \mc{J}_{[mode,q]}}\{y\in \R^{p(N-2\nu+1)}:\|y_{J}\|_2 \leq t\}$ is a set defined by constraints on subvectors corresponding to structured index collections.
Here $\mc{J}_{[mode,q]}$ denotes the collection of index sets corresponding to the aggregation structure of the test, and $y_J$ is the subvector indexed by $J$.
For example, when $mode=\cdot, q=2$, $\mc{J}{[\cdot,2]}$ consists of subsets of size $s$ within each time block:
\begin{equation}\label{eq:jset}
\small
\mc{J}_{[\cdot,2]}=\cup \{J: |J|=s \text{ and } J \subset \{pn+1,\ldots,pn+p\} \text{ for some } n=0,\ldots,N-2\nu\}.
\end{equation}
The maximum test corresponds to the special case $s=1$.
For mode-specific test, additional  row or column constraints are imposed.
In all cases, the set $A$ belongs to the class of generalized sparsely convex sets. This representation allows us to apply the Gaussian and bootstrap approximation results developed above to conduct inference for the proposed tests.

We implement inference using a Gaussian multiplier bootstrap adapted to the change point setting. Let $(e_1^b,\ldots,e_N^b)\sim\mathcal{N}(0,\bm{\Theta})$ be multiplier sequences independent of the data, where $\bm{\Theta}$ is defined using a kernel function as in Section~\ref{sec:approximation}.
To account for potential structural breaks, we adapt the bootstrap procedure in Section \ref{sec:approximation} by using piecewise centering based on pre- and post-change sample means, i.e., 
$\bar{\X}_n^{-}=n^{-1} \sum_{i=1}^n \X_i$ and $\bar{\X}_n^{+}=(N-n)^{-1} \sum_{i=n+1}^N \X_i$.
The bootstrap CUSUM statistic is then given by
\begin{equation}\label{eq:cnb}
\small
		\mbf{C}_{n}^b(\X)
		=  \sqrt{\frac{n(N-n)}{N}} \left(\frac{1}{N-n}\sum_{i=n+1}^N e_i^b \left(\X_{i}-\bar{\X}_n^{+}\right)-
		\frac{1}{n}\sum_{i=1}^n e_i^b \left(\X_{i}-\bar{\X}_n^{-}\right)\right),
\end{equation}
and the corresponding test statistic is
\begin{equation}\label{eq:tmodeqb}
	T_{[mode,q]}^b(\X)=\max_{\nu \leq n \leq N-\nu}\|\mbf{C}_{n}^b(\X)\|_{[mode,q]}.
\end{equation}
The p-value $\hat{P}_{[mode,q]}$ is calculated by (\ref{eq:pmode2}).

For the adaptive test in Section \ref{sec:adtest}, the test statistic is defined as the minimum of multiple p-values $T_{\ad}(\X)=\min_{mode \in \{1,2,\cdot\}, \, q \in \{2,\infty\}}\hat{P}_{[mode,q]}$.
A natural approach to approximate its distribution is the double bootstrap \citep{hall1986bootstrap,beran1988prepivoting}, but this is computationally expensive.
Moreover, since the p-values $\hat{P}_{[mode,q]}$ themselves are approximated by multiplier bootstrap, it is not clear how to conduct the second-level bootstrap based on (\ref{eq:cnb}).
Therefore, we propose a parallel bootstrap procedure by generating an additional independent set of bootstrap samples to approximate the reference distribution of the first-stage bootstrap statistics. Specifically,
generate $(e_1^b, \ldots, e_N^b) \sim \mc{N}(0, \bm{\Theta})$  for $b=B+1,\ldots,2B$, independently of $\X_1,\ldots, \X_N$ and and the first-stage multipliers $(e_1^{b}, \ldots, e_N^{b})$ for $b=1,\ldots,B$.
Obtain  $T_{[mode,q]}^{b}(\X)$ for $b=B+1,\ldots,2B$ by (\ref{eq:Tmodeq}) as a reference distribution. 
Specifically, the $b$th bootstrap analogue of the adaptive statistic is then defined as
\begin{align}\label{eq:tadb}
	T_{\ad}^b(\X)&=\min_{mode \in \{1,2,\cdot\}, \,q \in \{2,\infty\}}\hat{P}_{[mode,q]}^b, \text{ with }\\
	\hat{P}_{[mode,q]}^b&=\frac{1}{B}\sum_{b'=B+1}^{2B} \mbb{I}\left(T_{[mode,q]}^{b'}(\X) > T_{[mode,q]}^b(\X)\right).\nonumber
\end{align}
This construction avoids nested resampling by using a second, independent set of bootstrap samples, and we therefore refer to it as the parallel bootstrap.
This method is computationally efficient  and applicable to multiplier bootstrap.
The p-value of $T_{\ad}(\X)$ is then estimated by
\begin{equation*}
	\hat{P}_{\ad}=\frac{1}{B}\sum_{b=1}^B \mbb{I}\left(T_{\ad}^b(\X) < T_{\ad}(\X)\right),
\end{equation*}
and we reject $H_0$ at level $\alpha$ if $\hat{P}_{\ad} \le \alpha$.

\begin{remark}[Other bootstrap methods for the adaptive test]\label{remark:bootstrap}
    Alternative approaches for speeding up the double bootstrap, such as the fast double bootstrap \citep{davidson2007improving} 
    have been proposed to reduce computational burden. However, these methods are not directly applicable in our multiplier bootstrap setting.
    Additional theoretical and empirical comparisons  with the low-cost bootstrap \citep{zhou2018unified} are provided in the Supplementary Material Section \ref{sec:boot}. Empirically, the proposed parallel bootstrap achieves accurate size control with manageable computational cost.
\end{remark}

\section{Theoretical properties}\label{sec:theory}

In this section, we establish the theoretical properties of the proposed tests, including size validity and power analysis. We analyze the mode-specific test $T_{1,2}(\X)$, $T_{2,2}(\X)$ and the adaptive test $T_{\ad}(\X)$, which are shown in Sections  \ref{sec:individual-theory} and \ref{sec:adaptive-theory}, respectively.
The results build on the Gaussian and bootstrap approximations developed in Section~\ref{sec:GA}.

We impose the following regularity conditions on the error process ${\Err_i}$ in model~(\ref{eq:model}).
Similar to (\ref{eq:mixing}), define the $\alpha$-mixing coefficient for $\{\Err_i\}$ as $\alpha_N(k)$.
Let $b>0$ be a constant and let $D_N\geq 1$ be a sequence of constants which could diverge to infinity as $N$ increases.
Note that the $[mode,q]$ norm uses at most $p_{\max} \equiv \max(p_1,p_2)$ elements in each CUSUM matrix.

\begin{assumption}\label{ass:mixing}
	For any $k\geq 1$, $
	\alpha_N(k) \leq K_1 \exp(-K_2 k^{\gamma}),
	$ where $K_1>0 ,K_2>0, \gamma>0$ are some universal constants independent of $N$.
\end{assumption}

\begin{assumption}\label{ass:E}
  $\E\left[\exp\left(\left|\Err_{ij_1j_2}\right|/D_N\right)\right]\leq 2$ for all $i=1,\ldots,N$, $j_1=1,\ldots, p_1$ and $j_2=1,\ldots, p_2$.
\end{assumption}

\begin{assumption}\label{ass:M}
	$N^{-1} \Var[\sum_{i=1}^{N}(\bm{v}^{\top}\vect(\Err_i))] \geq b$ for some universal constant $b>0$ and all $\bm{v} \in \mbb{S}^{p-1}$ with $\|\bm{v}\|_0 \leq p_{\max}$.
\end{assumption}

\begin{assumption}\label{ass:kernel}
	The Kernel function $\mc{K}(\cdot)$ is continuously differentiable with bounded derivatives on $\R$ and satisfies (i) $\mc{K}(0)=1$, (ii) symmetric, and (iii) $|\mc{K}(x)| \lesssim |x|^{-\vartheta}$ as $|x| \to \infty$ for some constant $\vartheta>1$.
	Assume the bandwidth $l_N \asymp \nu^{\iota}$ with $0<\iota<(\vartheta-1)/(3\vartheta-2)$.
\end{assumption}

The first three assumptions correspond respectively to  Conditions \ref{mixing}, \ref{espa} and \ref{mspad}  used in the Gaussian approximation theories in Section \ref{sec:spa}.
Assumption \ref{ass:mixing} imposes geometric $\alpha$-mixing, while Assumption~\ref{ass:E} requires sub-exponential tails, which can be relaxed to polynomial moment conditions as discussed earlier.
Assumption \ref{ass:M} ensures a non-degeneracy condition for projections. 
Assumption \ref{ass:kernel} specifies standard conditions on the kernel function and bandwidth used in the bootstrap procedure.

\subsection{Theoretical results for the mode-specific test statistics}\label{sec:individual-theory}
In this section, we derive the theoretical properties of the mode-specific test $T_{1,2}(\X)$ and $T_{2,2}(\X)$, regarding their size and power.
We first establish the bounds on the Kolmogorov distance between $T_{[mode,q]}(\X)$ and $T_{[mode,q]}^b(\X)$ with $mode \in \{1,2\}$ and $q=2$, which is defined as
\begin{equation*}
\rho\left(T_{[mode,q]}(\X),T_{[mode,q]}^b(\X)\right)=\sup_{t \in \R}\left|\Pro\left(T_{[mode,q]}(\X)\leq t\right)-\Pro\left(T_{[mode,q]}^b(\X)\leq t | \X\right)\right|.
\end{equation*}
Henceforth, we set $s=p_2$ if $mode=1$, and $s=p_1$ if $mode=2$.

\begin{theorem}\label{thm:size}
  Under (\ref{eq:model}) and $H_0$, suppose Assumptions  \ref{ass:mixing}-\ref{ass:kernel} hold.
  Assume $p \geq N^{\tau}$ for some universal constant $\tau>0$, and  $(s\log p)^{3-\gamma}=o(N^{\gamma/3})$.
  Define
  \begin{align}\label{eq:eta}
  	\eta_N (s) \equiv & \frac{N^{2/9}D_{N}^{2/3}s^{(1+4\gamma)/(3\gamma)}(\log p)^{(1+2\gamma)/(3\gamma)}}{\nu^{1/3}}
  	+\frac{N^{7/18}D_{N}s^{13/6}(\log p)^{7/6}}{\nu^{1/2}}\nonumber\\
  	& + \frac{N^{1/3}  D_{N}^{2/3} s(\log p)^{h_2}}{\nu^{h_1}},
  \end{align}
  where $1/3<h_1\leq (\iota+1)/3$ is a constant depending on $(\iota,\vartheta)$, and $h_2>2/3$ is a constant depending on $(\gamma,\vartheta)$.
  Then, for some constant $h_3>0$, there exists a constant $h_4>0$ depending on $(b,h_3)$ such that with probability at least $1-p^{-h_3}$ we have
  \begin{equation*}
    \rho\left(T_{[mode,2]}(\X),T_{[mode,2]}^b(\X)\right) \leq
    h_4 \eta_N (s).
  \end{equation*}
\end{theorem}

For the approximation error rate $\eta_N(s)$, although the expression may appear complex at first glance, all three terms share the same  form,
$N^{c_1}D_N^{c_2} s^{c_3} (\log p)^{c_4} \nu^{-c_5}$ with positive constants $c_1,c_2,c_3,c_4,c_5$ satisfying $c_5>c_1$. The first two terms have exponents determined by the $\alpha$-mixing decay rate and arise from the Gaussian approximation step in the proof. The third term, which comes from the bootstrap approximation, has exponents that depend on the $\alpha$-mixing decay rate, the kernel decay rate, and the bandwidth.

Theorem \ref{thm:size} immediately implies uniform size validity of the proposed tests.
\begin{corollary}\label{coro}
  Under all assumptions in Theorem \ref{thm:size}, we have
  \[
  \sup_{\alpha \in (0,1)}\left|\Pro\left(
  \hat{P}_{[mode,2]}\leq \alpha
  \right)
  -\alpha\right|\lesssim \eta_N (s)+\exp\{-2\eta_N^2 (s)B\}.
  \]
  Consequently, if $B$ is sufficiently large and
  \begin{equation}\label{eq:etanb}
  	\max \left(
  	N^{\frac{2}{3}} D_{N}^{2} s^{\frac{1+4\gamma}{\gamma}} (\log p)^{\frac{1+2\gamma}{\gamma}},
  	\;
  	N^{\frac{7}{9}} D_{N}^2 s^{\frac{13}{3}} (\log p)^{\frac{7}{3}},
  	\;
  	\left\{ N^{\frac{1}{3}} D_{N}^{\frac{2}{3}} s(\log p)^{h_2} \right\}^{\frac{1}{h_1}}
  	\right)
  	= o(\nu),
  \end{equation}
  then, $\Pro\left(
  \hat{P}_{[mode,2]}\leq \alpha
  \right) \to \alpha$ uniformly in $\alpha \in (0,1)$ as $N \to \infty$.
\end{corollary}

\begin{remark}[Choices of boundary removal]
  The form of $\eta_N(s)$ and (\ref{eq:etanb}) cast some light on  the impact of $\nu$ regarding the validity of the test size.
  For simplicity, consider the case of a constant $D_N$, in which case a trade-off comes into play when selecting $\nu$: a larger $\nu$ improves approximation accuracy and hence size control, but reduces the search range and may miss change points near the boundaries.
  For instance, we may set $\nu=c_0N$ with a constant $0<c_0<1$, i.e., at a  linear rate of $N$, which is commonly adopted in the change point  literature for multivariate time series  \citep[Table 3]{wang2023computationally}.
  
\end{remark}

Next, we study the power of the tests by characterizing the signal strength required to distinguish $H_1$ from $H_0$.
While Assumption \ref{ass:E} ensures size validity, power analysis requires additional control on the joint behavior of the noise under aggregation. We therefore impose a stronger condition Assumption \ref{ass:Ep} that guarantees sub-exponential behavior for linear projections of rows and columns.
Denote $[\Err_i]_{j:}$ as the $j$-th row in the $\Err_{i}$ matrix at time $i$, and $[\Err_i]_{:j}$ as the $j$-th column at $i$.

\begin{assumptionprime}{2}\label{ass:Ep}
	(i)
	$\E\left[\exp\left(\left|\bm{v}^{\top}[\Err_i]_{j:}\right|/D_N\right)\right]\leq 2$ for all $1\leq i \leq N$, $1 \leq j \leq p_1$ and  $\bm{v} \in \mbb{S}^{p_2-1}$.
	(ii) $\E\left[\exp\left(\left|\bm{v}^{\top}[\Err_i]_{:j}\right|/D_N\right)\right]\leq 2$ for all $1\leq i \leq N$, $1 \leq j \leq p_2$ and  $\bm{v} \in \mbb{S}^{p_1-1}$.
\end{assumptionprime}

The next theorem establishes the power results for the case $mode=1$. The corresponding results for for $mode=2$ follow analogously.
WLOG, henceforth assume that $ \|\bs{\delta}_N\|_{F} \lesssim D_N/\sqrt{l_N}$, which ensures the validity of the bootstrap approximation under the alternative.
Stronger signals only improve power and lead to trivial detection.

\begin{theorem}\label{thm:power}
  Under (\ref{eq:model}) and $H_1$, suppose $\nu \leq u \leq N-\nu$ and 
  that Assumptions \ref{ass:mixing}, \ref{ass:Ep}, \ref{ass:M} and \ref{ass:kernel} hold.
  Assume $p \geq N^{\tau}$ for some universal constant $\tau>0$, and  $(s\log p)^{3-\gamma}=o(N^{\gamma/3})$ with $s=p_2$.
  Suppose the mean shift signal magnitude is large enough such that 
  \begin{equation}\label{eq:delta}
    \left\|\bs{\delta}_N\right\|_{[1,2]}  
    \geq c_1D_N\frac{\log^{1/2}(N p_13^{p_2})+ \log^{1/2}(2/\alpha)}{\{u(N-u)/N\}^{1/2}},
  \end{equation}
  where $c_1>0$ is a constant depending only on $b$.
  Then, there exists a constant $c_{2}>0$ depending only on $b$ such that
  \begin{equation*}
    \Pro
    \left(
    \hat{P}_{[1,2]}\leq \alpha
    \right)
    \geq 1-c_2\eta_N(s)-2\exp(-\alpha^2B/2).
  \end{equation*}
  Consequently, if $B$ is sufficiently large and (\ref{eq:etanb}) holds, we have $\Pro
  \left(
  \hat{P}_{[1,2]}\leq \alpha
  \right) \to 1$ as $N \to \infty$.
\end{theorem}

\subsection{Theoretical results for adaptive test statistics}\label{sec:adaptive-theory}
We now study the size and power properties of the matrix adaptive test $T_{\ad}(\X)$, which combines multiple individual statistics. The goal is to show that the adaptive procedure maintains valid size while automatically adapting to different structural forms of change.
Define $\eta_N \equiv \eta_N(p_{\max})$.

\begin{theorem}\label{thm:size-ad}
  Under all assumptions in Theorem \ref{thm:size}, we have 
  \[
  \sup_{\alpha \in (0,1)}\left|\Pro(\hat P_{\ad} \leq \alpha)-\alpha\right|\lesssim \eta_N+\exp(-2\eta_N^2B).
  \]
  Thus, if $B$ is sufficiently large and (\ref{eq:etanb}) holds with $s$ replaced by $p_{\max}$, we have $\Pro(\hat P_{\ad} \leq \alpha)\to\alpha$ uniformly for $\alpha \in (0,1)$ as $N \to \infty$.
\end{theorem}

To study power, we impose a stronger tail condition that controls all sparse linear projections of the noise. This assumption strengthens Assumptions \ref{ass:E} and \ref{ass:Ep} and ensures that aggregation across different modes remains well behaved.
\begin{assumptionpprime}{2}\label{ass:Epp}
	$\E\left[\exp\left(\left|\bm{v}^{\top}\vect(\Err_i)\right|/D_N\right)\right]\leq 2$ for all $i=1,\ldots,N$ and  $\bm{v} \in \mbb{S}^{p-1}$ with $\|\bm{v}\|_0 \leq p_{\max}$.
\end{assumptionpprime}

The next result establishes the power of the adaptive test. It shows that the test is consistent as long as the signal is detectable by at least one of the component statistics.

\begin{theorem}\label{thm:power-ad}
    Under (\ref{eq:model}) and  $H_1$, suppose $\nu \leq u \leq N-\nu$ and 
  that Assumptions \ref{ass:mixing}, \ref{ass:Epp}, \ref{ass:M} and \ref{ass:kernel} hold.
  Assume $p \geq N^{\tau}$ for some universal constant $\tau>0$, and  $(p_{\max}\log p)^{3-\gamma}=o(N^{\gamma/3})$.
  Suppose the mean shift signal magnitude is large enough such that  there exist $mode\in \{ 1,2,\cdot\}$ and $q\in \{2,\infty\}$ such that
  \begin{equation}\label{eq:delta-ad}
\left\|\bs{\delta}_N\right\|_{[mode,q]}  
    \gtrsim D_N\frac{\log^{1/2}(\lambda_{[\text{mode}, q]})+ \log^{1/2}(8/\alpha)}{\{u(N-u)/N\}^{1/2}},
 \end{equation}
 with
\[
\lambda_{[\text{mode}, q]} = \left\{
\begin{array}{ll}
	Np & \text{if } q = \infty, \\
	N\left(\dfrac{p}{s}\right)^s 
	& \text{if } \text{mode} = \cdot,\, q = 2, \\
	N p_1 \cdot 3^{p_2} & \text{if } \text{mode} = 1,\, q = 2,\\
	N p_2 \cdot 3^{p_1} & \text{if } \text{mode} = 2,\, q = 2.
\end{array}
\right.
\]
  Then, if $B$ is sufficiently large and (\ref{eq:etanb}) holds with $s$ replaced by $p_{\max}$, we have $\Pro\left(\hat P_{\ad} \leq \alpha\right) \to 1$ as $N \to \infty$.
\end{theorem}
\begin{remark}[Power of the adaptive test]
	The theoretical conditions of $\lambda_{[mode,q]}$ established in Theorems \ref{thm:power-ad} indicate that different test statistics target different structural patterns of change.
	The max test with $q=\infty$ is driven by the largest coordinate and is therefore suited for extremely sparse changes. Compared to the max test, the condition for the mode-free test with $mode=\cdot,q=2$ is weaker for alternatives in which several components change simultaneously. Compared to these two tests, the conditions of mode-specific tests with $mode=1,2,q=2$ are weaker for alternatives in which the change is concentrated within certain rows or columns, illustrating its advantage in detecting structured mode-sparse changes.
	Theorem \ref{thm:power-ad} shows that the proposed adaptive test is consistent whenever at least one of the above conditions holds. Consequently, the adaptive procedure inherits the detection capability of all these tests and provides a robust solution that is effective across a broad range of change-point structures.
\end{remark}

\begin{remark}[Adaptive versus mode-specific tests in theory]\label{rem:vs-theory}
Comparing the adaptive and mode-specific tests reveals a tradeoff between robustness and efficiency. The adaptive test acts as a robust device against model uncertainty, at the cost of local efficiency and rate sharpness.
For size, it involves a larger effective dimension through $p_{\max} = \max(p_1,p_2)$, which may lead to more restrictive rate conditions, especially when $p_1$ and $p_2$ are imbalanced. For power, a stronger condition (Assumption \ref{ass:Epp}) is required, and combining multiple tests introduces a multiplicity penalty in the signal strength through $\log^{1/2}(2m/\alpha)$, where $m=4$ is the number of individual tests.
By contrast, mode-specific tests achieve sharper rates when the change aligns with the corresponding structure, but may lose power under misspecification.
\end{remark}

\section{Simulation}\label{sec:simu}

In this section, we report the empirical performance of the proposed  tests in terms of  size and power, and conduct a comparative analysis against several existing state-of-the-art methods for multivariate data.
Table \ref{tab:method} summarizes these methods, including their dependence settings (column I/D, where I denotes methods designed for independent data and D denotes methods allow dependent data), aggregation procedures and  approaches for obtaining p-values, and the value of $\gamma$.
Note that we use the prefix `M' and `V' to denote whether these methods are designed for matrices or vectors.
We refer to our mode-specific tests in (\ref{eq:T}) as M-mode1 for $mode=1$ and M-mode2 for $mode=2$, and we refer our adaptive test in (\ref{eq:Tad}) as M-adapt.

\begin{table}[!htbp]
  \caption{Methods for simulation}
  \centering
  \small
  \begin{tabular}[t]{lllll}
    \toprule
    Method & I/D & Aggregation &Calibration  & Reference \\
    \midrule
    M-adapt  & D & adaptive & Multiplier bootstrap  & Our paper\\
    M-mode1  & D & $[1,2]$ norm & Multiplier bootstrap  & Our paper\\
    M-mode2  & D & $[2,2]$ norm & Multiplier bootstrap  & Our paper\\
     V-DC  & D & Double CUSUM & Bootstrap  & \cite{cho2016changepointa}\\
    V-YC  & I & $l_{\infty}$ norm & Multiplier bootstrap  & \cite{yu2021finite}\\
    V-LZZL & I & $[s_0,p]$ norm & Multiplier bootstrap  & \cite{liu2020unified}\\
    V-DMS0 & I & adaptive & Asymptotic  & \cite{wang2023computationally}\\
    V-DMS05 & I & adaptive & Asymptotic  & \cite{wang2023computationally}\\
    \bottomrule
  \end{tabular}
\label{tab:method}
\end{table}

\subsection{Empirical issues}\label{sec:empirical}

The implementation of the proposed tests requires the specification of  the boundary removal parameter $\nu$. 
Theorem \ref{thm:size} provides some theoretical guidance on how to determine $\nu$, which  is related  to the series length $N$ and the dimension $p$.
A common recommendation in the literature is to set $\nu=\lfloor0.2N \rfloor$ \citep[Table 3]{wang2023computationally}.
In our simulations, we tried choices around $0.2N$.
Specifically, we set $\nu=60$ when $N=250$, $\nu=80$ when $N=500$, and $\nu=150$ when $N=750$.
In all numerical work,  the data were first re-scaled by their mean absolute deviation series-by-series before applying a test for detecting a mean change. 

For the mode-free test in the adaptive test, we use  $s=\sqrt{p}$, chosen to be comparable to the number of rows or columns in the mode-specific tests. 
Other choices, including alternative deterministic choices or data-driven selection of $s$, could also be considered.
A systematic investigation of this choice is beyond the scope of the present paper and we leave it for future work.

For the bootstrap procedure,  we set the number of bootstraps $B=400$. We adopt the kernel in 
\cite{andrews1991heteroskedasticity}, i.e., the quadratic spectral kernel
\begin{equation*}
	\mathcal{K}_{\rm QS}(x)=\frac{25}{12\pi^{2}x^{2}}\bigg\{\frac{\sin(6\pi x/5)}{6\pi x/5}-\cos(6\pi x/5) \bigg\}\,.
\end{equation*} 
This kernel function guarantees the positive definiteness of the long-run covariance matrix estimate.
Also, we adapt his data-driven bandwidth selection procedure by fitting an AR(1) model to each component series of the  matrix time series and setting the bandwidth
%
 $l=1.3221(\hat{a}N)^{1/5}$ with
$
\hat{a}=\{\sum_{j=1}^p4\hat{\rho}_j^2\hat{\sigma}_j^4(1-\hat{\rho}_j)^{-8}\}/\{\sum_{j=1}^p\hat{\sigma}_j^4(1-\hat{\rho}_j)^{-4}\}$, where 
 $\hat{\rho}_j$ and $\hat{\sigma}_j^2$ are the estimated autoregressive coefficient and innovation variance of
 the $j$-th component series of the (vectorized) matrix time series, respectively.

Although we focus on change point testing, the change point location could also be estimated after $H_0$ is rejected.
For the mode-specific test $T_{[1,2]}$ and $T_{[2,2]}$, it is natural to use the following estimator $$\hat{u}_{[mode,2]}=\arg\max_{\nu \leq n \leq N-\nu}\|\mbf{C}_{n}(\X)\|_{[mode,2]}.$$
For the adaptive test, we propose to use the estimator with the smallest p-value, i.e., 
\begin{align}
    \label{eq:cpestimator}
    \begin{split}        \hat{u}_{\ad}&=\arg\max_{\nu \leq n \leq N-\nu}\|\mbf{C}_{n}(\X)\|_{[mode^*,q^*]}, \\
  [mode^*,q^*] &=\arg\min_{mode \in \{1,2,\cdot\}, \, q \in \{2,\infty\}}\hat{P}_{[mode,q]}.
    \end{split}
\end{align}

To handle  multiple change points, we combine our test with the binary segmentation method.
Take the adaptive test $T_{\ad}(\X)$ and estimator $\hat{u}_{\ad}$ as examples.
Specifically, within the interval $(s,e)$, if $H_0$ is rejected by $T_{\ad}(\X)$, then we estimate the change point by (\ref{eq:cpestimator}), i.e.,  $u=\arg\max_{s+\nu \leq n \leq e -\nu}\|\mbf{C}_{n}(\X)\|_{[mode^*,q^*]}$.
Subsequently, we divide the interval into two sub-intervals $(s,u)$ and $(u,e)$, and repeat the aforementioned procedure within each sub-interval. 
This iterative process continues until no further change points are detected.
In Supplementary Section \ref{sec:simuesti}, we show simulation results for change point estimation.

\subsection{Size performance}
For assessing size performance, we generate data from model (\ref{eq:model}) with $\bs{\delta}_N=0$ under different scenarios of serial length, matrix dimensions, covariance structures, and temporal dependence. Specifically, we consider serial lengths $N=250, 500,$ and $750$, and matrix dimensions $p_1=5, p_2=10$ or $p_1=p_2=20$.
Regarding the covariance structure and dependence, the error matrices $\Err$ are generated from an element-wise autoregressive process: 
\begin{equation}\label{eq:ar}
    \Err_{ij_1j_2}=\phi\Err_{i-1j_1j_2}+\mathbf{V}_{ij_1j_2},
\end{equation}
where $\mathbf{V}$ is a matrix-valued innovation process such that $\vect(\mathbf{V})$ are
independent and identically Gaussian distributed with zero mean and  covariance matrix $\Sigma=\Cov(\vect(\mathbf{V}))$ taking one of the  following structures:
\begin{itemize}[topsep=0pt, itemsep=0pt]
  \item Cov 1 (Uncorrelated): $\Sigma$ is an identity matrix.
  \item Cov 2 (Row-column separability): $\Sigma=\Sigma_c \otimes \Sigma_r$, where $\Sigma_c$ is randomly generated according to $\Sigma_c=\mathbf{Q} \Lambda \mathbf{Q}^{\top}$, where the eigenvalues in the diagonal matrix $\Lambda$ are the absolute values of i.i.d. standard normal random variates, and the eigenvector matrix $\mathbf{Q}$ is a random orthonormal matrix. $\Sigma_r$ is similarly generated. 
  \item Cov 3 (Banded): $\Sigma$ is such that 
  $\Cov({\varepsilon}_{j_1k_1},{\varepsilon}_{j_2k_2})=0.5^{|j_1-j_2|}*0.3^{|k_1-k_2|}$. 
  \item Cov 4 (Compound symmetry): the diagonal elements of $\Sigma$ equal 1 while the non-diagonal elements equal 0.2.
\end{itemize}
The autoregressive coefficient is set as $\phi=0.2$ or $0$. The former represents the dependent setting for which the method is designed, while the latter corresponds to the independence case, included to demonstrate that the proposed procedure remains valid when independence arises as a special case of the dependence framework.

\renewcommand{\arraystretch}{0.8}
\begin{table}
\centering
\caption{Empirical size results for tests of nominal size $0.05$}
\centering
\resizebox{\ifdim\width>\linewidth\linewidth\else\width\fi}{!}{
\begin{tabular}[t]{lrrrrrrrr}
\toprule
  & M-adapt & M-mode1 & M-mode2 & V-DC & V-YC & V-Liu & V-DMS0 & V-DMS05\\
\midrule
\multicolumn{9}{l}{\textbf{$\phi=0,N=250,p_1=5, p_2=10$}}\\
\hspace{1em}Cov1 & 0.041 & 0.027 & 0.044 & 0.014 & 0.048 & 0.068 & 0.090 & 0.085\\
\hspace{1em}Cov2 & 0.056 & 0.047 & 0.052 & 0.028 & 0.057 & 0.063 & 0.085 & 0.080\\
\hspace{1em}Cov3 & 0.050 & 0.036 & 0.048 & 0.030 & 0.052 & 0.069 & 0.081 & 0.079\\
\hspace{1em}Cov4 & 0.051 & 0.050 & 0.051 & 0.021 & 0.049 & 0.068 & 0.081 & 0.074\\
\multicolumn{9}{l}{\textbf{$\phi=0.2,N=250,p_1=5, p_2=10$}}\\
\hspace{1em}Cov1 & 0.044 & 0.039 & 0.047 & 0.047 & 0.381 & 0.691 & 1.000 & 1.000\\
\hspace{1em}Cov2 & 0.063 & 0.058 & 0.061 & 0.071 & 0.344 & 0.493 & 0.983 & 0.986\\
\hspace{1em}Cov3 & 0.061 & 0.045 & 0.061 & 0.068 & 0.373 & 0.522 & 0.994 & 0.994\\
\hspace{1em}Cov4 & 0.060 & 0.064 & 0.060 & 0.081 & 0.349 & 0.410 & 0.978 & 0.981\\
\multicolumn{9}{l}{\textbf{$\phi=0.2,N=500,p_1=5, p_2=10$}}\\
\hspace{1em}Cov1 & 0.038 & 0.031 & 0.035 & 0.038 & 0.444 & 0.756 & 1.000 & 1.000\\
\hspace{1em}Cov2 & 0.053 & 0.052 & 0.050 & 0.058 & 0.397 & 0.504 & 0.980 & 0.982\\
\hspace{1em}Cov3 & 0.044 & 0.039 & 0.071 & 0.051 & 0.400 & 0.552 & 0.996 & 0.995\\
\hspace{1em}Cov4 & 0.057 & 0.050 & 0.053 & 0.049 & 0.389 & 0.374 & 0.986 & 0.994\\
\multicolumn{9}{l}{\textbf{$\phi=0.2,N=750,p_1=20, p_2=20$}}\\
\hspace{1em}Cov1 & 0.030 & 0.027 & 0.029 & 0.039 & 0.696 & 1.000 & 1.000 & 1.000\\
\hspace{1em}Cov2 & 0.030 & 0.031 & 0.037 & 0.025 & 0.654 & 0.989 & 1.000 & 1.000\\
\hspace{1em}Cov3 & 0.033 & 0.043 & 0.029 & 0.033 & 0.672 & 0.998 & 1.000 & 1.000\\
\hspace{1em}Cov4 & 0.068 & 0.060 & 0.061 & 0.021 & 0.605 & 0.535 & 1.000 & 1.000\\
\bottomrule
\end{tabular}}
\label{tab:size}
\end{table}

Table \ref{tab:size} reports the empirical sizes of all competing tests at the nominal level $0.05$, based on $1000$ Monte Carlo replications for each scenario. Overall, the proposed matrix-based tests (M-adapt, M-mode1, and M-mode2) exhibit stable and accurate size control across all scenarios considered. Their empirical rejection rates remain close to the nominal level under both independent ($\phi=0$) and dependent ($\phi=0.2$) settings, and are largely insensitive to variations in sample size and covariance structure. When the matrix dimension increases to $(p_1,p_2)=(20,20)$ under dependence ($\phi=0.2$), the proposed tests become slightly conservative, likely due to the increased effective dimension of the aggregated statistics. Nevertheless, the empirical sizes remain stable across the scenarios considered.

Among the competing procedures, the vector-based method V-DC, which is designed for dependent data, also maintains reasonably good size control, although it tends to be somewhat conservative in several scenarios. In contrast, the methods developed for independent data (V-YC, V-LZZL, V-DMS0, and V-DMS05) behave as expected when the observations are independent ($\phi=0$), but exhibit substantial size distortions once temporal dependence is present. In particular, under $\phi=0.2$, these procedures frequently over-reject the null hypothesis, with empirical sizes increasing dramatically as the sample size or matrix dimension grows.
These results highlight the importance of explicitly accounting for matrix structure and temporal dependence in high-dimensional change-point testing.

\subsection{Power performance}

\begin{figure}[t]
  \centering
  \subfigure{\includegraphics[width=0.25\textwidth]{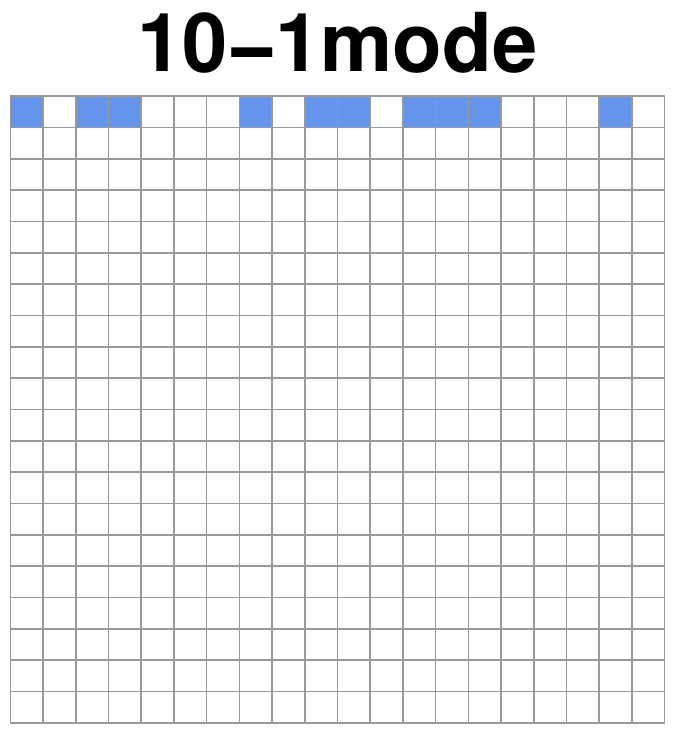}}
  \subfigure{\includegraphics[width=0.25\textwidth]{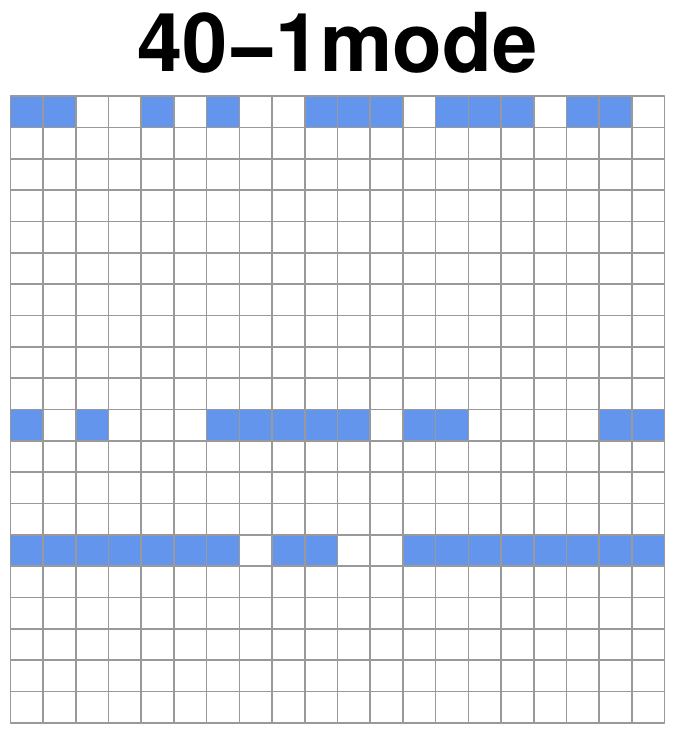}}
  \subfigure{\includegraphics[width=0.25\textwidth]{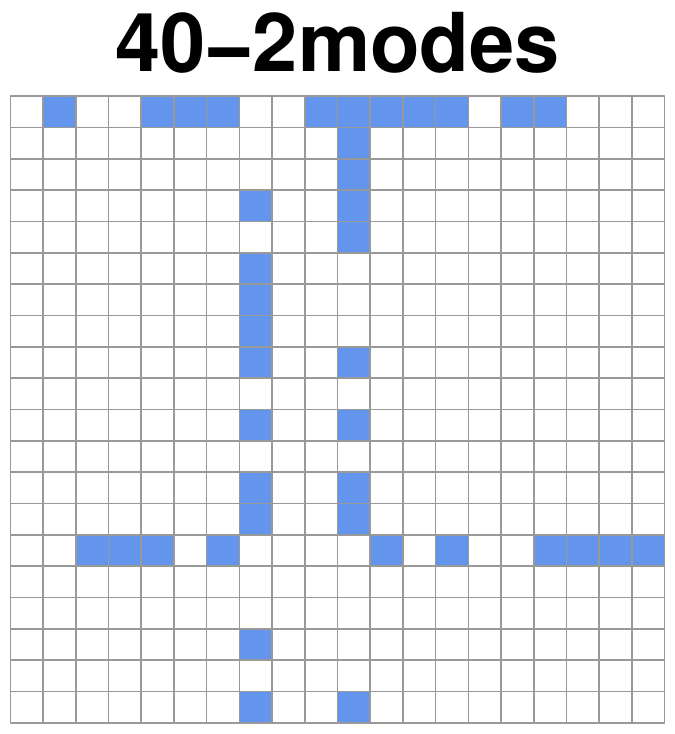}}\\
  \subfigure{\includegraphics[width=0.25\textwidth]{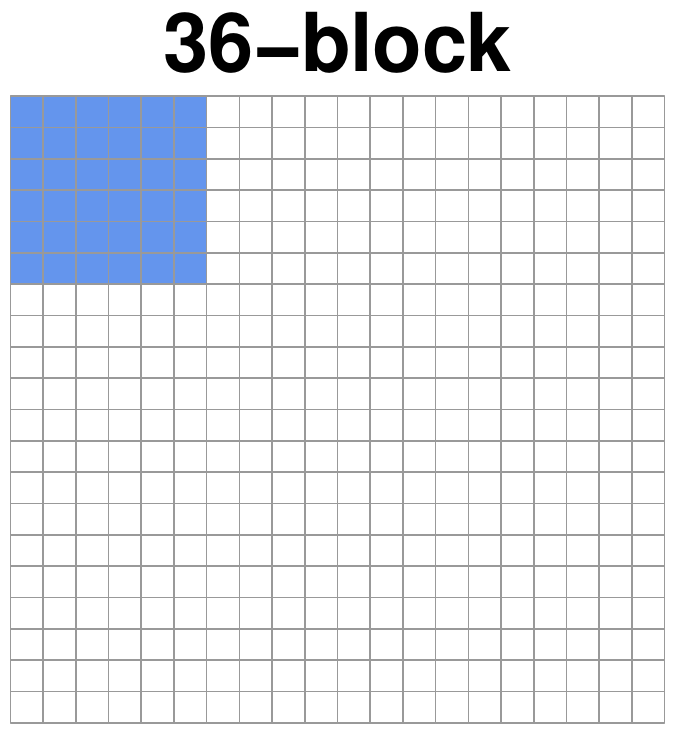}}
  \subfigure{\includegraphics[width=0.25\textwidth]{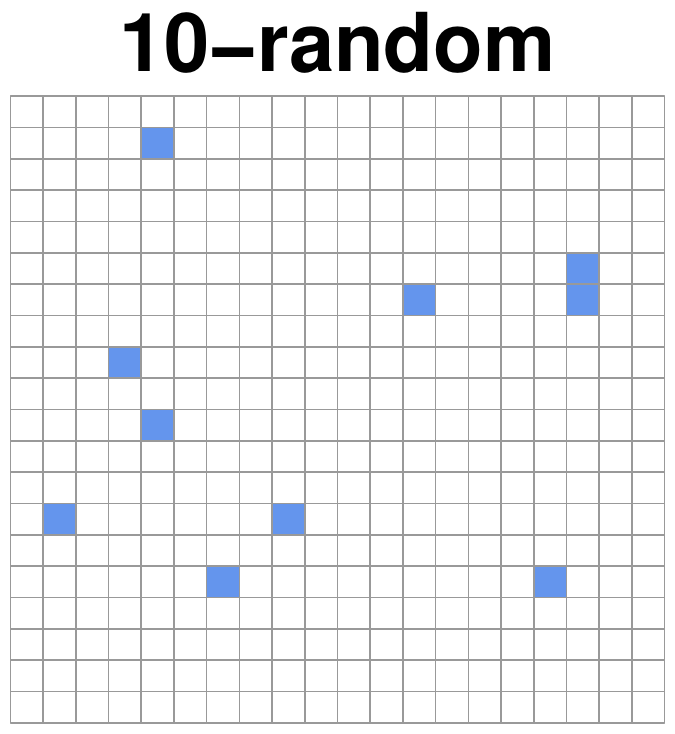}}
  \subfigure{\includegraphics[width=0.25\textwidth]{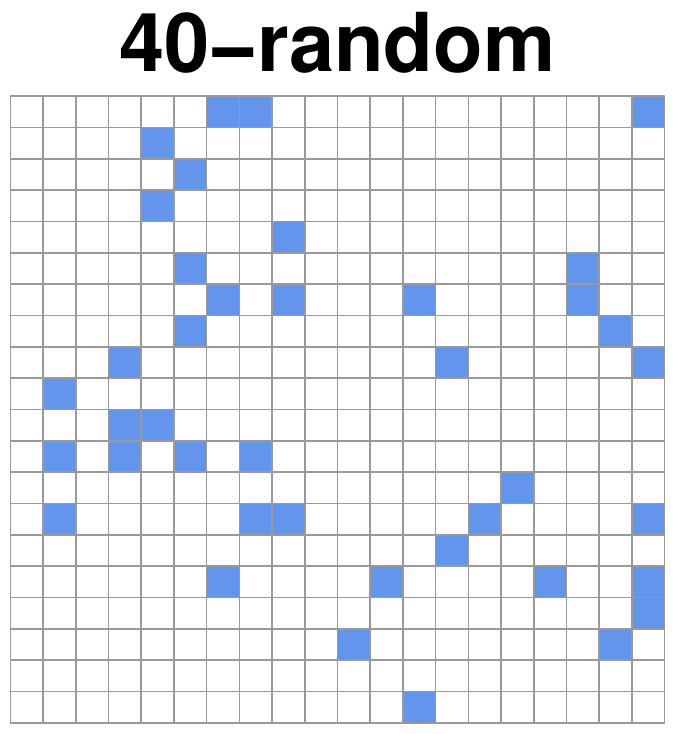}}
  \caption{Scenarios of $\bs{\delta}_N$ for detection. Blue elements are nonzero and identical.}
  \label{fig:alignment}
\end{figure}

Next, we assess the power of the proposed tests, under different scenarios of temporal dependence (AR coefficient $\phi$), series length ($N$), change-point location ($u$), break alignments ($\bs{\delta}_N$) and sparsity.
Specifically, we generate matrix time series from (\ref{eq:model}) with $p_1=p_2=20$.
The errors $\Err_i$ follow the AR model (\ref{eq:ar}) with coefficient $\phi$, and the innovation are Gaussian distribution with zero mean and covariance matrix 
$\Sigma=\Cov(\vect(\mathbf{V}))$ as in the setting Cov 4.
For the first three arguments, we consider four settings:(i) $\phi=0.2,N=500,u=125$; (ii) $\phi=0.2,N=250,u=125$; (iii) $\phi=0,N=500,u=125$; (iv) $\phi=0,N=250,u=125$.
Note that in settings (i) and (iii), the change point is located at the first quarter of the series, whereas in settings (ii) and (iv) it is located at the center of the series.
We consider six settings of break alignment in the matrix $\bs{\delta}_N$. See Figure \ref{fig:alignment}, where all $k$ non-zero elements are painted blue and they are  identically equal to  $a>0$.
We  assess the power performance against a sequence of $a$ values.
The alignment could be matrix-featured (1mode, 2modes, block) or vector-based (random), and the sparsity ranges from highly sparse ($k=10$) to relatively dense ($k=36$ and $k=40$).

\begin{figure}[!htbp]
  \centering
  \includegraphics[width=0.95\textwidth]{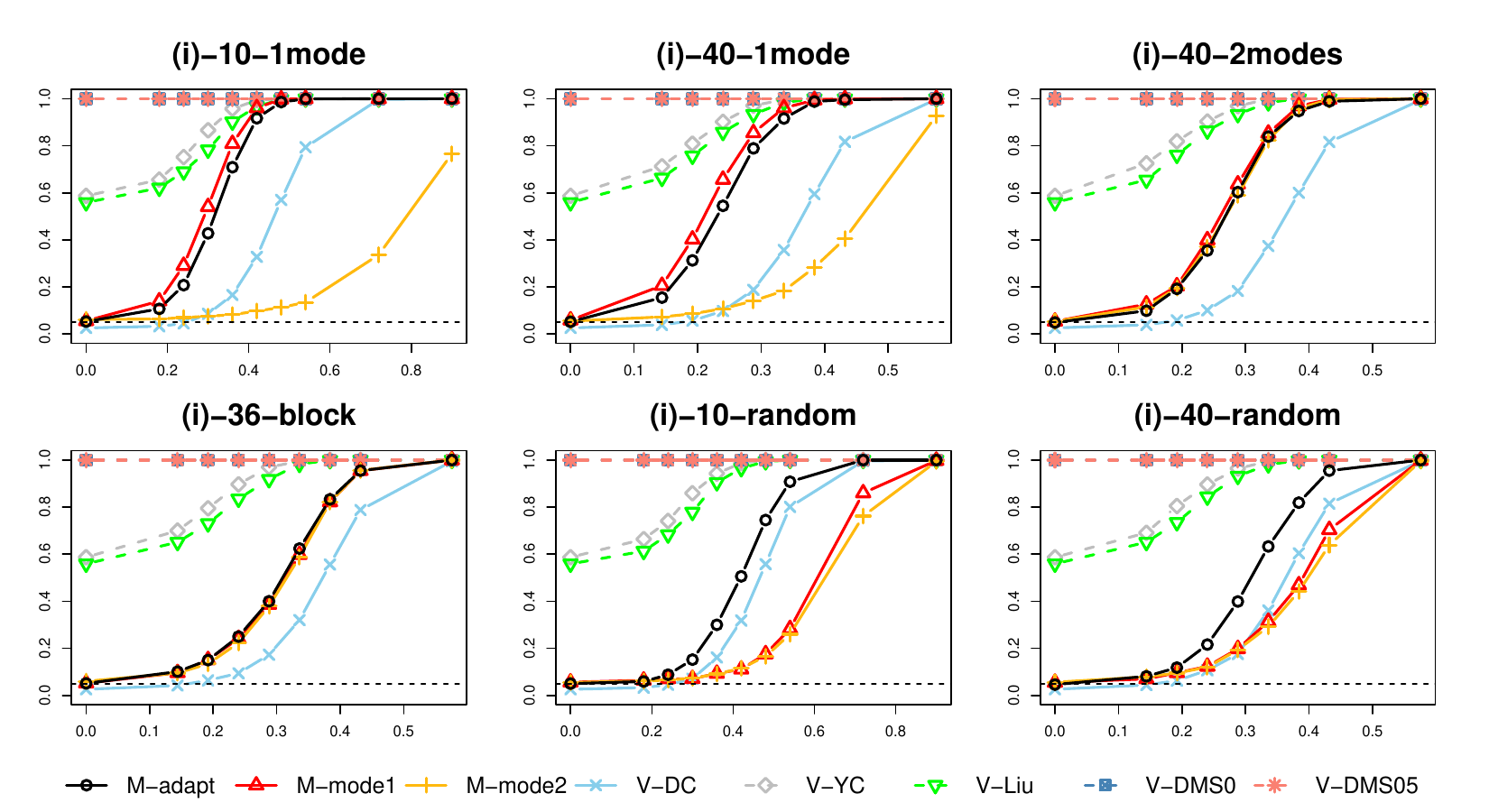}
  \includegraphics[width=0.95\textwidth]{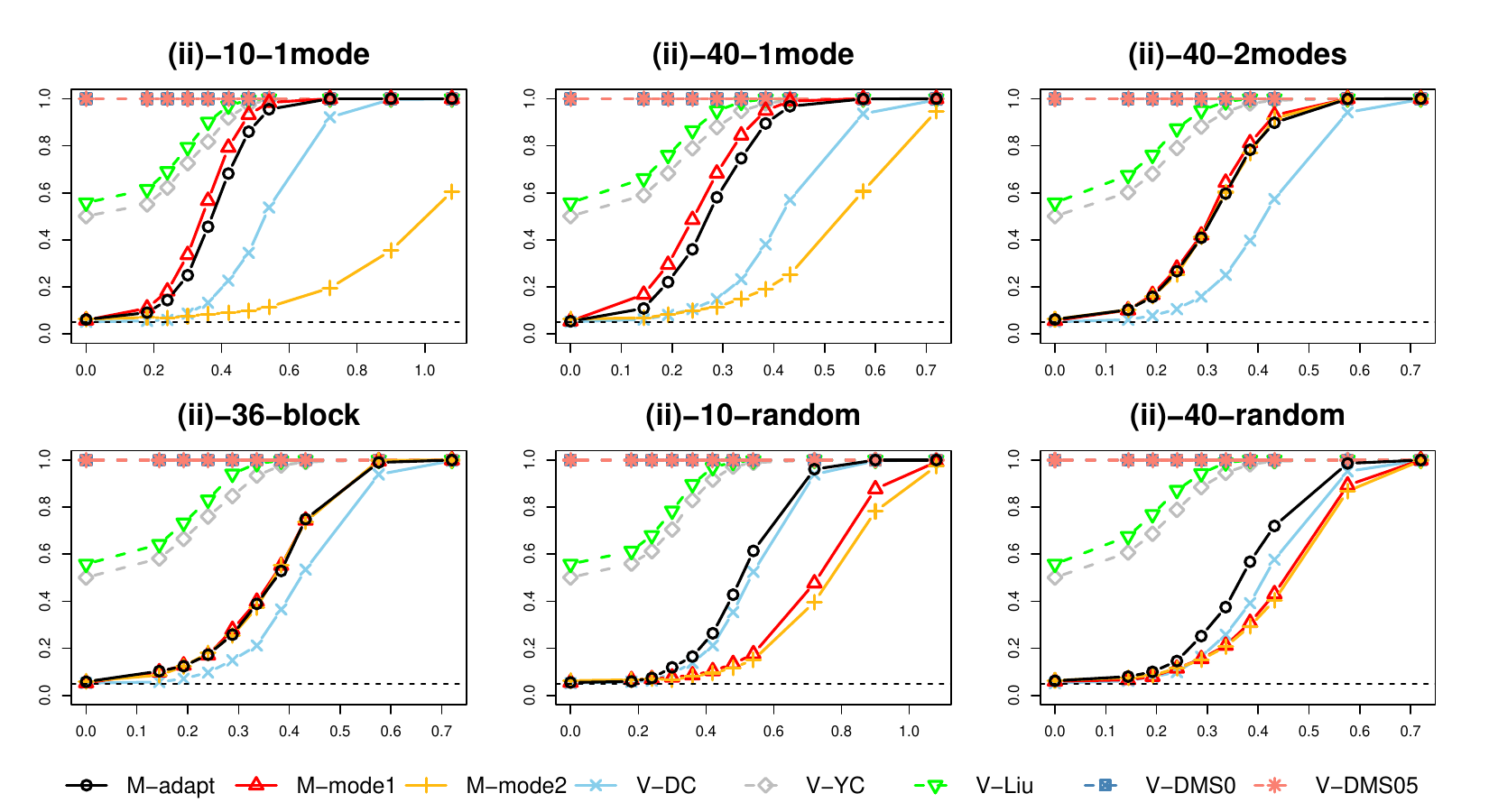}
  \includegraphics[width=0.95\textwidth]{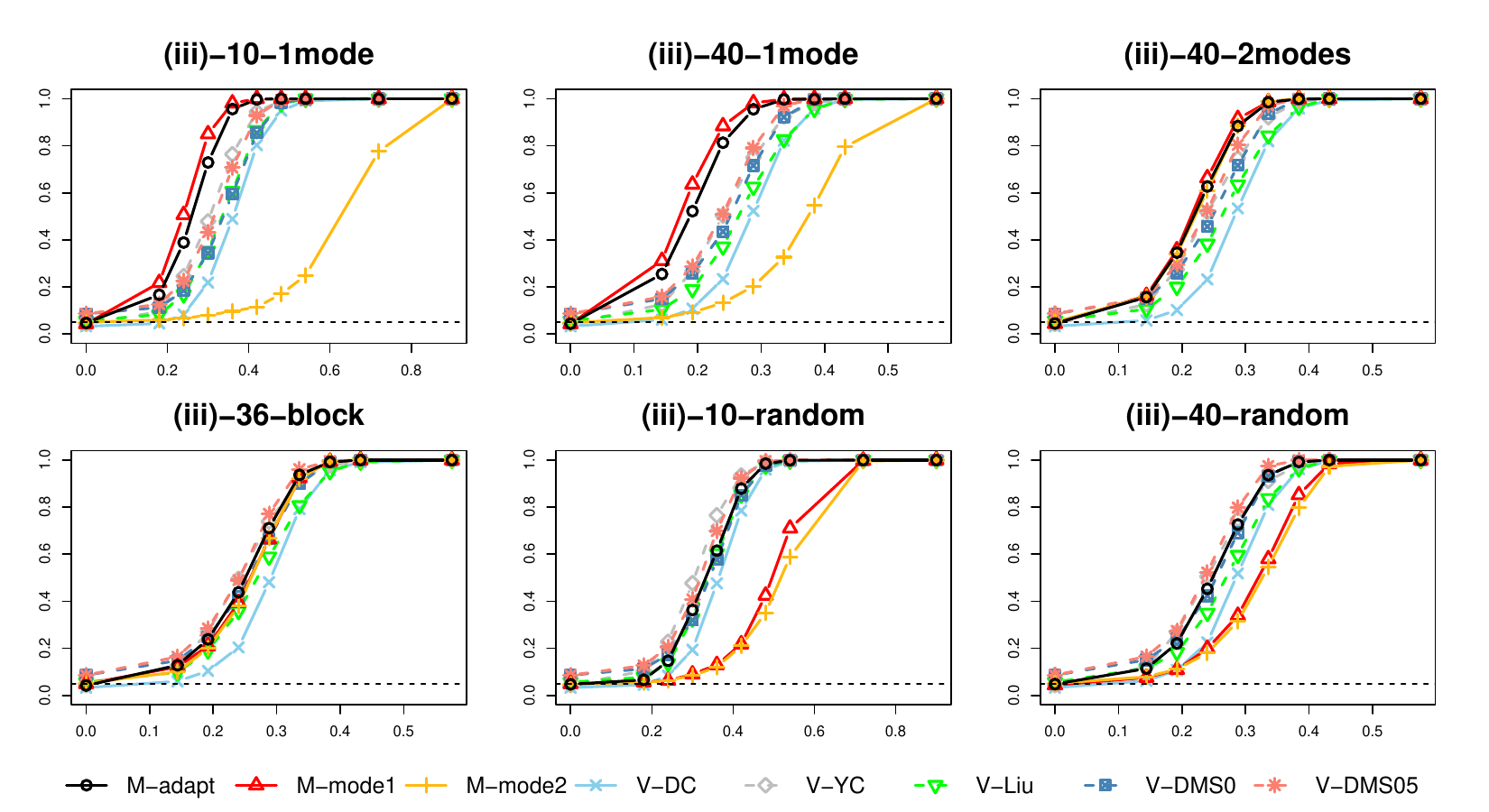}
  \caption{ Power results under different scenarios: (i) $\phi=0.2,N=500,u=125$; (ii) $\phi=0.2,N=250,u=125$; (iii) $\phi=0,N=500,u=125$.}
  \label{fig:power}
\end{figure}


Figure \ref{fig:power} reports the empirical power curves as a function of the change magnitude $a$ for scenarios (i)–(iii), based on 1000 Monte Carlo replications for each setting; the results for scenario (iv) are provided in the \emph{Supplementary Materials} Section \ref{sec:powermore}.

We first compare the results with respect to temporal dependence. In scenarios (i) and (ii), where the data exhibit serial dependence ($\phi=0.2$), the methods designed for independent observations suffer from substantial size distortions (see Table \ref{tab:size}), and therefore their power curves are not directly comparable. In this case, the most relevant benchmark is the dependent-data procedure V-DC. Across most settings, the proposed matrix-based tests achieve comparable or higher power than V-DC and tend to detect moderate changes earlier as the signal magnitude increases. In scenario (iii), where the observations are independent ($\phi=0$), the independent-data methods become properly calibrated and achieve competitive power. In this setting, the proposed tests remain comparable with these methods.

We then examine the effect of break alignment and sparsity. When the change pattern follows a matrix structure (1mode, 2modes, and block), clear differences emerge among the aggregation strategies. In the 1mode setting, where the change occurs along rows, M-mode1 consistently achieves higher power than M-mode2, since M-mode1 aggregates row-wise signals whereas M-mode2 aggregates column-wise signals. This illustrates that directional aggregation can be highly effective when the break pattern aligns with the targeted matrix dimension, but may become less sensitive when the alignment is mismatched. The adaptive test M-adapt performs robustly across these configurations and typically achieves power close to the best-performing directional test. For the random alignment cases, where the signal does not exhibit a clear matrix structure, the performance differences between matrix-based and vector-based procedures become smaller, and the advantage of directional aggregation diminishes. In particular, both M-mode1 and M-mode2 lose power relative to the structured settings. Nevertheless, M-adapt remains competitive across alignments and sparsity levels, reflecting its ability to adapt to different signal patterns.
Finally, the methods exhibit stable performance across different series lengths and change-point locations.

Overall, the proposed mode-specific tests can yield substantial gains when the change pattern aligns with the matrix structure, but may lose power when the alignment is absent, whereas the adaptive test provides reliable performance across different levels of temporal dependence, break alignment, and change-point location.

\section{Application}\label{sec:app}

We illustrate the proposed tests with the New York yellow taxi trip data that is maintained by the NYC Taxi and Limousine Commission (TLC) and is publicly available at \url{https://www.nyc.gov/site/tlc}. 
This dataset provides a high-resolution view of urban transportation activity and is widely used to study demand dynamics and operational efficiency in large metropolitan systems.

Our objective is to detect structural changes in the spatial–temporal pattern of taxi demand. Such changes are of direct relevance for transportation management and urban operations, as they reflect shifts in travel behavior, demand intensity, and system-wide usage patterns. Identifying these changes can inform resource allocation, pricing strategies, and policy evaluation in transportation systems.

\begin{figure}[!htbp]
  \centering
  \includegraphics[width=0.45\textwidth,height=0.45\textwidth]{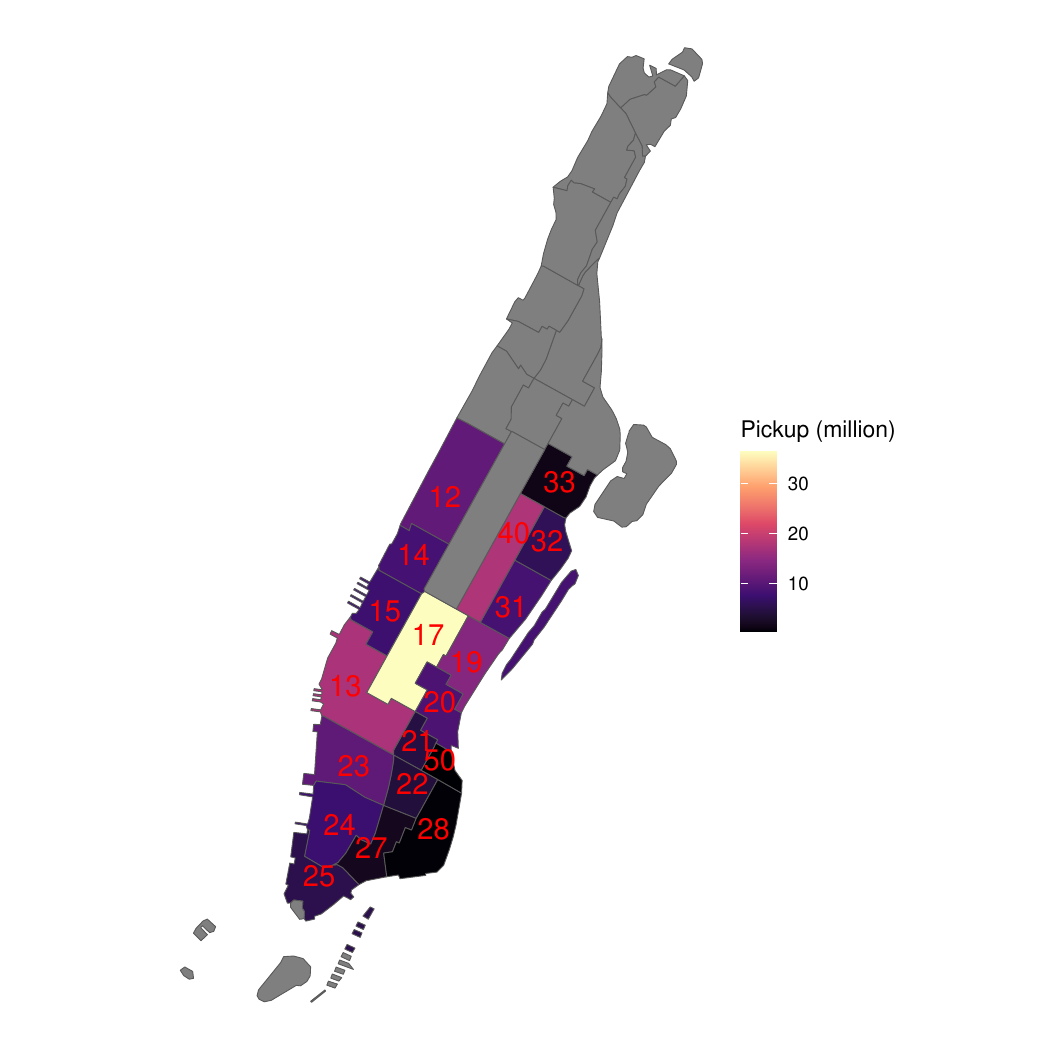}
  \includegraphics[width=0.4\textwidth,height=0.45\textwidth]{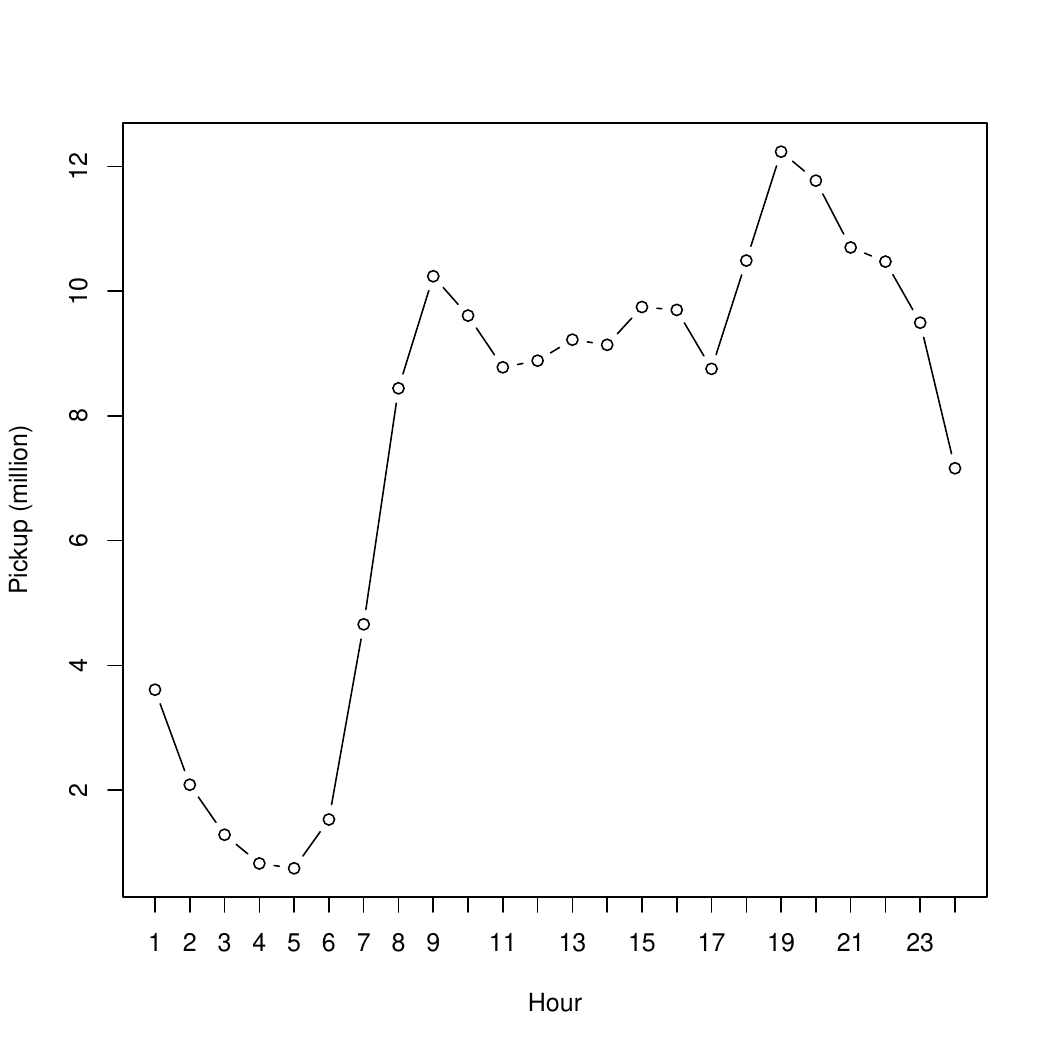}
  \caption{Taxi pickups in study zones (Left) with NTA labels and  by hours (Right).}
  \label{fig:nyc}
\end{figure}

We focus on Manhattan areas where most taxi activity is concentrated (lower, middle, upper west, and upper east), and aggregate trips by Neighborhood Tabulation Areas (NTAs).
See Figure \ref{fig:nyc} for the map of the study area.
There are 19 NTAs in total, and their labels start with the prefix `MN' followed by a two-digit code.
As the traffic pattern differ profoundly between business  and non-business days, to reduce heterogeneity, we remove data collected on non-business days and treat the data as regularly spaced time series.
Thus, over the study period 2017 - 2019, the data constitute a matrix time series $\X= \{\X_1,\ldots,\X_N \}$ with $\X_t \in \R^{p_1 \times p_2}$, where $p_1=19$, $p_2=24$ and $N=754$.
Note that $\X_{ijt}$ denotes the number of taxi trips picked up from the $i$th region (NTA) within the $j$th hour on the $t$th day.
Figure \ref{fig:cusum}  displays the matrix of univariate time series plots in 2017 (only data within 1-8 am are shown due to space limit), each of which is scaled by their median absolute deviation, and their corresponding  CUSUM curves defined by (\ref{eq:cn}).
Figure \ref{fig:nyc} summarizes the marginal patterns across the two modes, i.e., $\sum_{j,t}\X_{ijt}$ for the pick-up location pattern, and $\sum_{i,t}\X_{ijt}$ for the pick-up hour pattern.

\begin{figure}[h]
  \centering
  \includegraphics[width=0.8\textwidth]{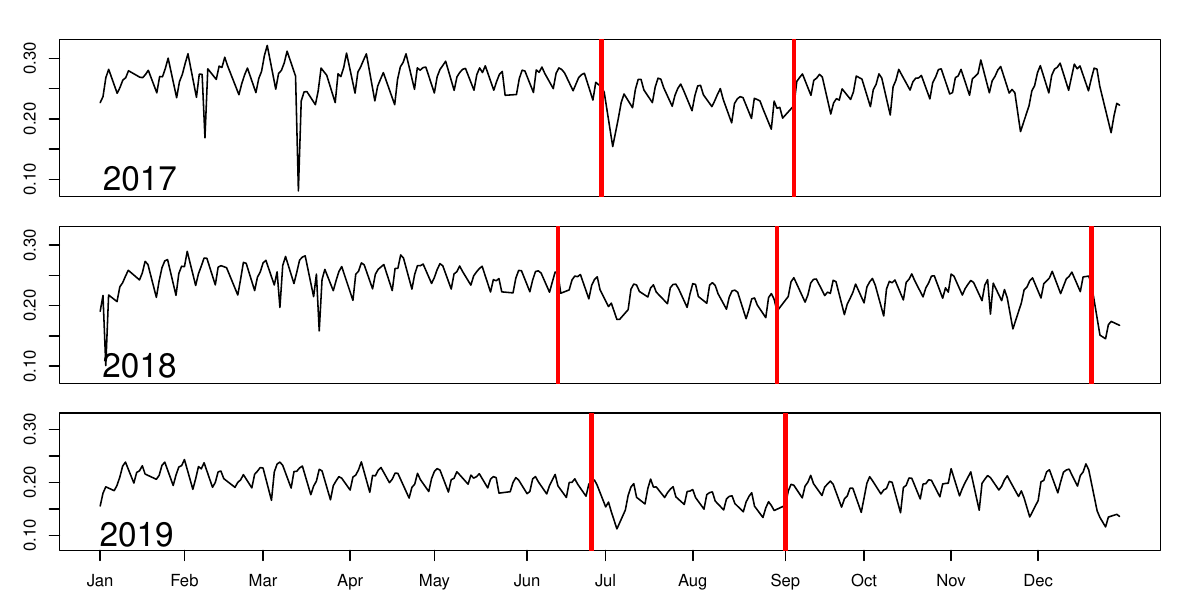}
  \caption{Change points detected by M-adapt (red lines); the daily aggregated taxi pickup pattern (in millions) is shown for reference.}
  \label{fig:nyccp}
\end{figure}

Figure \ref{fig:nyccp} shows the estimated change-point locations (red lines) obtained from the proposed M-adapt test with boundary removal $\nu=30$ and bootstrap number $B=1000$.
The daily aggregated pick-up pattern, i.e., $\sum_{i,j}\X_{ijt}$, is plotted only as a reference to facilitate visualization.
To better appreciate the within- and between-year patterns, the data are plotted year by year, on three parallel panels.
Our methodology detects a total of seven change points: two in 2017, three in 2018, and two in 2019. Several of these breaks exhibit a recurring seasonal pattern, particularly those occurring in late June and early September, which coincide with the beginning and end of the academic summer break, during which Manhattan’s taxi demand typically declines.
These findings align with the results in \cite{zhou2022ringcpd}, who identified similar breaks in the 2014 trip record counts for the New York Central Park region.
In addition to these seasonal effects, a distinct structural break is observed in December 2018, likely reflecting both holiday-related demand changes and regulatory adjustments affecting High-Volume For-Hire Vehicles (HVFHV).

\begin{figure}[h] \includegraphics[width=0.9\textwidth,height=0.25\textheight]{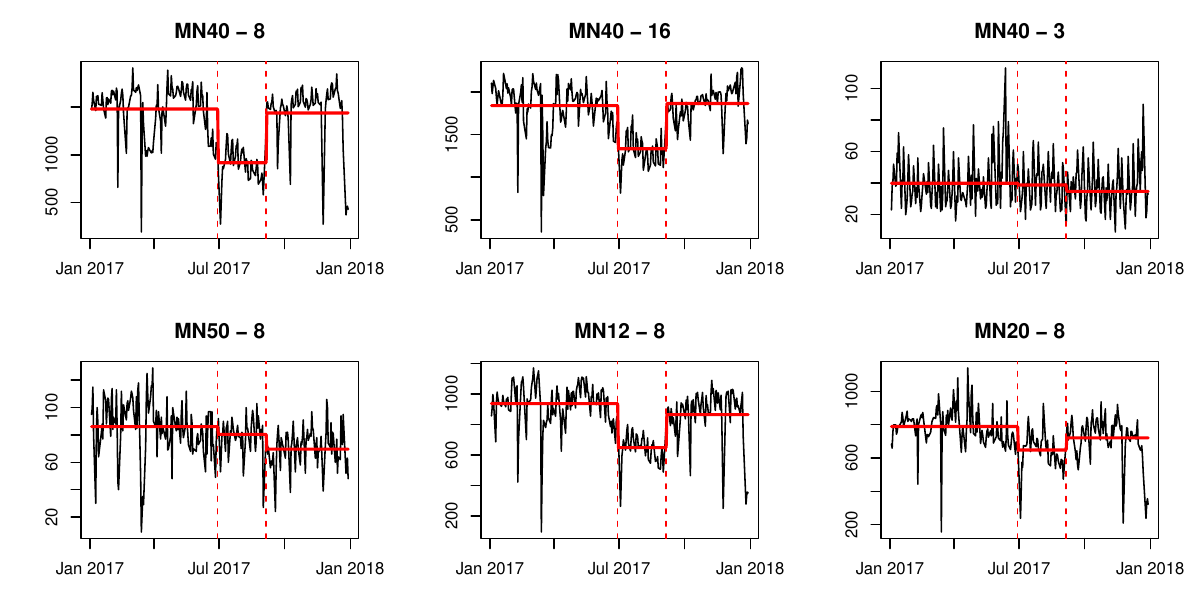}
  \caption{Examples of taxi pickups in certain `region - hour' pairs. Red lines are the segment means delineated  by the matrix change points (red dash lines).}
  \label{fig:nyccpeg}
\end{figure}

Some of the detected change points are not visually prominent in the aggregated series shown in Figure \ref{fig:nyccp}, as the break signals may arise from heterogeneous and possibly sparse changes in individual components of the matrix. To further illustrate this, Figure \ref{fig:nyccpeg} presents selected time series of taxi pickups at specific locations and hours in 2017.
The figure shows that the structural changes can manifest differently across components: in some cases, multiple series exhibit coordinated shifts, while in others the changes are more localized. 
This pattern highlights a key feature of the data: structural changes occur along specific dimensions rather than uniformly across the system. As a result, methods based on vectorization may fail to detect such changes, whereas the proposed matrix-based approach is able to capture both structured and sparse shifts.

Overall, the detected change points indicate that structural shifts in NYC taxi demand are often system-wide but driven by complex patterns across locations and time.
These shifts reflect both recurring behavioral dynamics and exogenous factors such as policy changes. The results demonstrate that the proposed method can effectively identify such changes in high-dimensional settings, providing a useful tool for analyzing structural instability in urban transportation systems.


\section{Discussion}\label{sec:dis}

This paper studies the problem of change point detection for matrix-variate time series.
The key observation is that the break signals may conform to some mode-specific alignment, and thus mode-specific aggregation is proposed to enhance change point detection.
We complement this approach with vector-based aggregation, and 
form an adaptive test that remains effective across a wide range of signal configurations.
The parallel bootstrap method is proposed for calibrating the p-value of the adaptive test statistic, which is more  computationally efficient than the double bootstrap.
On the theoretical side, we establish size validity and power properties under temporal dependence and high dimensionality. In doing so, we derive refined Gaussian and bootstrap approximation bounds of sparsely convex sets for dependent data.
These bounds are  of independent interest for high dimension problems especially with divergent sparsity and dependence,
and thus serve as additional theoretical contributions.

Some future research directions: 
First, while the current framework focuses on mean changes, it would be of interest to extend the methodology to other types of matrix-valued data, such as covariance matrices or network adjacency matrices, where structural breaks may reflect changes in dependence or connectivity.
Second, our matrix method could be naturally extended to higher-order tensors.
In such settings, mode-specific aggregation can be defined along each tensor dimension, but the interaction across multiple modes introduces additional complexity in both modeling and inference. 
Third,
the refined Gaussian and bootstrap approximation results may be useful in other high-dimensional settings where signals are concentrated within groups or clusters of variables, such as submatrix-based comparisons of covariance structures and best subgroup detection problems.
Finally, extending the proposed framework to incorporate general $l_q$-type aggregation, thereby allowing for finer adaptivity to different sparsity regimes, is also an interesting direction.
We leave all these interesting topics as future work.

\section*{Supplementary Material}
The \emph{Supplementary Materials} includes all proofs and additional numerical results.




\bibliographystyle{abbrvnat}%
\bibliography{library.bib}

\end{document}